\documentclass[aps,prb,twocolumn, showpacs,floats]{revtex4}
\usepackage{bm}
\usepackage{graphicx}
\newcommand{\bi}[1]{{\boldsymbol{#1}}}

\begin{document}

\title{Moving and merging of Dirac points on a square lattice   and hidden symmetry protection  }
\author{Jing-Min Hou}\email{jmhou@seu.edu.cn}
\affiliation{Department of Physics, Southeast University, Nanjing,
211189, China}

\begin{abstract}
First, we study a square fermionic  lattice that supports the
existence of massless Dirac fermions, where the Dirac points are
protected by a hidden symmetry. We also consider two modified models
with a staggered potential and the diagonal hopping terms,
respectively. In the modified model with a staggered potential,  the
Dirac points exist in some range of magnitude of the staggered
potential, and   move  with the variation of  the staggered
potential. When the magnitude of the staggered potential reaches a
critical value, the two Dirac points merge. In the modified model
with the diagonal hopping terms, the Dirac points always exist and
just move   with the variation of  amplitude of the  diagonal
hopping. We develop a mapping method to find hidden symmetries
evolving with the parameters. In the two modified models, the Dirac
points are protected by this kind of hidden symmetry, and  their
moving
    and merging process can be explained by the   evolution of the  hidden symmetry
    along with the variation of the corresponding parameter.


\end{abstract}
\maketitle

\section{Introduction}

Success in the preparation of graphene has led to an enormous amount
of interests in massless Dirac fermions in condensed matter
physics.\cite{Novoselov1,Novoselov2,Zhang,Gusynin,Li} Many schemes
on the simulation of massless Dirac fermions in optical lattices
have been proposed theoretically
\cite{Hou2,Zhu,Goldman,Bercioux,Goldman2} and verified
experimentally.\cite{Tarruell} The recent discovery of topological
insulators\cite{Fu,Moore,Roy} and Weyl
semimetals\cite{Wan,Xu,Burkov,Jiang,Delplace} further facilitated
the research on massless Dirac fermions in condensed matter systems.

In condensed matter materials, Dirac fermions appear as emergent
particles near  Dirac points in the Brillouin zone, where the band
degeneracies occur.  Around Dirac points, the dispersion relation is
linear and can be described by the Dirac equation. Sometimes, the
chirality can be defined for massless Dirac fermions, which can be
considered as Weyl fermions.\cite{Hou1}   In three-dimensional
materials, the band degeneracy at Dirac points can be accidental.
The von Neumann-Wigner theorem tells us that,  to achieve such a
two-band accidental degeneracy, three real parameters are required
to be tuned.\cite{vNW} Thus, accidental band degeneracies are
vanishingly improbable in two dimensions if there are not additional
symmetry constraints.\cite{Balents} Therefore, in two dimensions,
the band degeneracy at Dirac points must be protected by some
symmetry. In general, the band degeneracy is protected by point
groups or time-reversal symmetry. Recently, the author has shown
that touching points of some two-dimensional lattices are protected
by a kind of hidden symmetry.\cite{Hou1}
  These hidden symmetries are discrete symmetries with antiunitary composite operators,
   which, in general,
  consist  of a translation, a complex conjugation and a sublattice exchange, and sometimes
  they also include a local gauge transformation and a rotation.

 In this paper, we  first consider a square fermionic lattice supporting the existence of massless Dirac fermions, where the Dirac points are protected by a hidden
 symmetry.  We regard this model as
 the original model.   We also consider two modified models with a staggered potential and the diagonal hopping terms, respectively.
In the modified model with a staggered potential, the two Dirac points with opposite topological
 charges move away from or towards  each other
  with increasing   magnitude of  the staggered  potential. When the magnitude of the staggered  potential arrives at a critical value,
   the two Dirac points with opposite topological charges merge, and if we continue to increase the magnitude of the staggered  potential,
   a gap opens and the system becomes an insulator.
   In the modified model with the diagonal hopping terms, the two Dirac points in the Brillouin zone move with increasing   amplitude of the diagonal hopping in
   two opposite directions, respectively. As the amplitude of the diagonal hopping approaches  infinity, the two Dirac points asymptotically approach the $k_x=0$ line,
   but they never vanish.
    We will show that, in these two modified models,  the Dirac points are protected by a hidden symmetry,
   the moving of Dirac points can be explained by the evolution of hidden symmetry along with the variation of the corresponding parameter, and
    the merging  of Dirac points in the modified model with a staggered potential  can be
    interpreted from the disappearance of the hidden-symmetry-invariant points in the Brillouin zone.

\begin{figure}[ht]
\includegraphics[width=0.48\columnwidth]{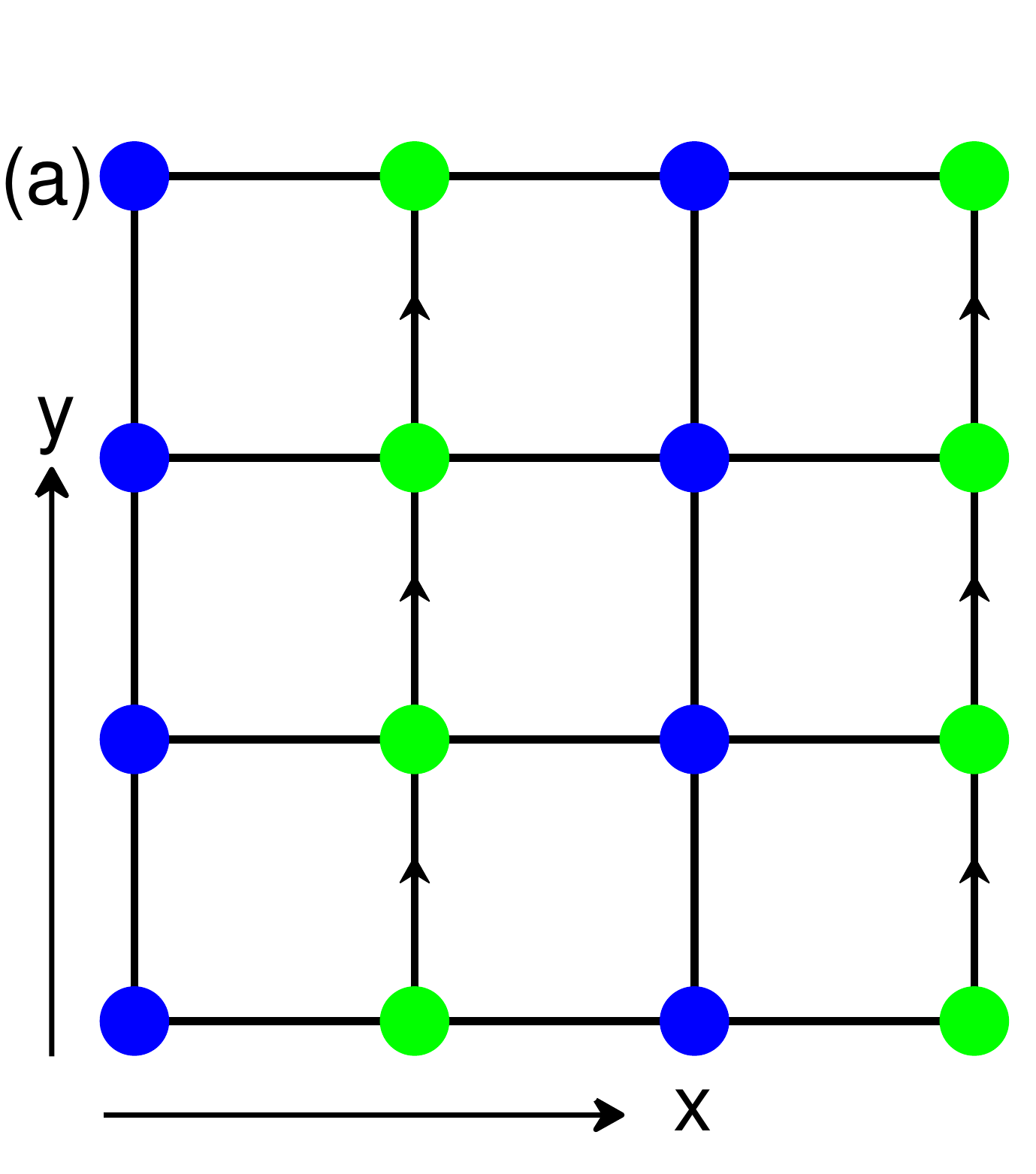}
\includegraphics[width=0.48\columnwidth]{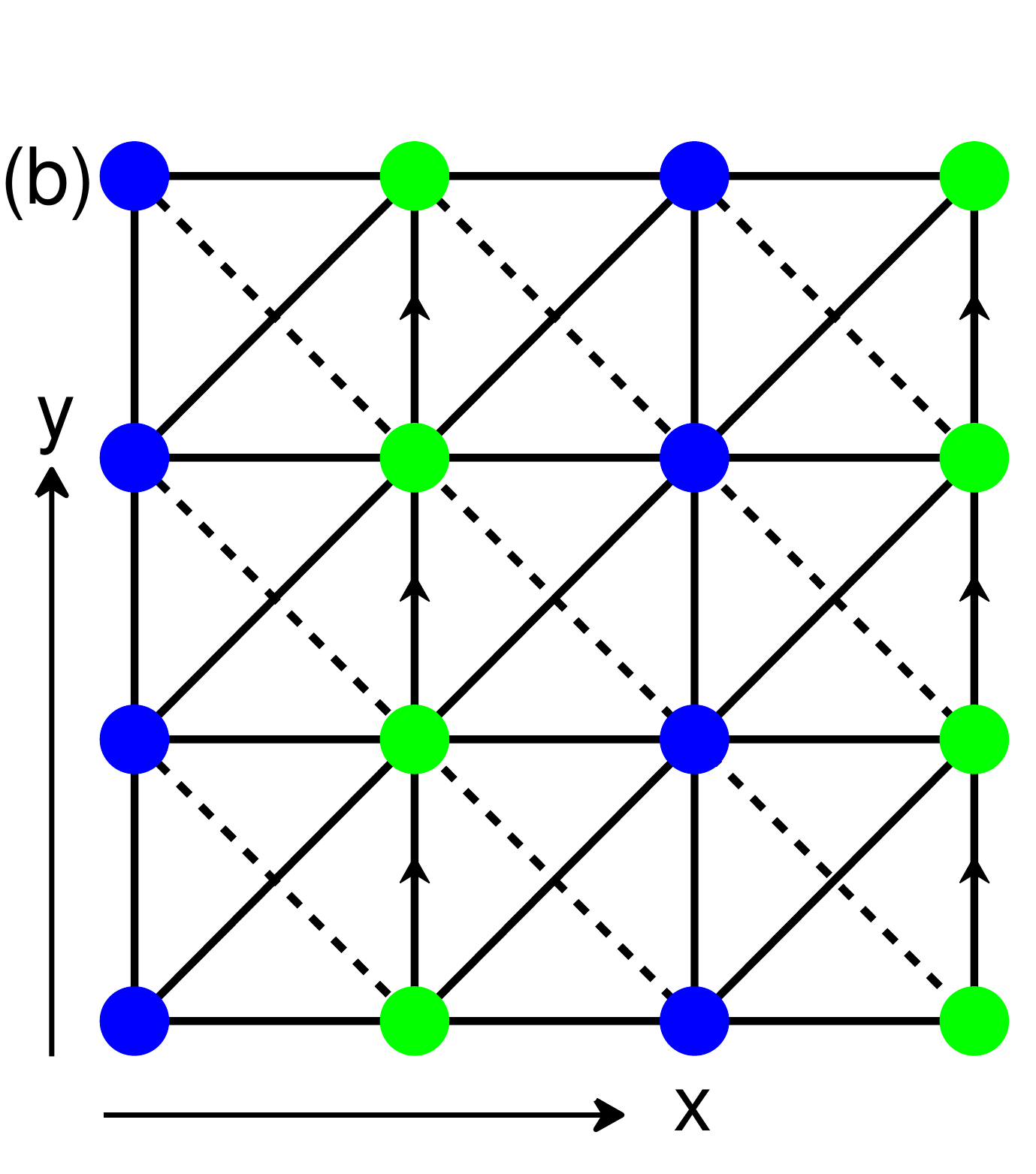}
\caption{(Color online). Schematic of the square lattice for (a) the
original model   and the modified model with a staggered potential,
(b) the modified model with the diagonal hopping terms. Here, the
arrows represent   a $\pi$ accompanying phase in the vertical
hopping; The dashed lines represent a $\pi$ accompanying phase  in
the diagonal hopping; the blue and green filled circles represent
the lattice sites of sublattices $A$ and $B$, respectively. }
\label{fig1}
\end{figure}

\section{Model}

First,  we consider the original model that consists of two
sublattices denoted as $A$ and $B$, respectively, as shown in
Fig.\ref{fig1}(a). The two sublattices have the lattice spacings $2d
$ and $d $ in the $x$ and $y$ directions, respectively. For
sublattice $B$, along the $y$ direction, there exists an
accompanying phase $\pi$ of hopping between two neighbor lattice
sites. For each sublattice, the primitive lattice vectors are
defined as $\bi{a}_1=(2d,0)$ and $\bi{a}_2=(0,d)$. In the following
process, for simplicity, we assume $d=1$.  The tight-binding
Hamiltonian for the original model can be written as,
\begin{eqnarray}
  H_0&=&-\sum_{i\in A} [ t_x  a^\dag_{i}  b_{i+\hat{x}} +
t_xa^\dag_{i}b_{i-\hat{x}} \nonumber\\&&+ t_y
  a^\dag_{i}
a_{i+\hat{y}}+ t_y e^{-i\pi}  b^\dag_{i+\hat{x}}
  b_{i+\hat{x}+\hat{y}}+
{\rm
H.c.}],\label{tbh}
\end{eqnarray}
 where $a_i  $ is the
annihilation operator that destructs  a particle in the Wannier
state $w^A_i(\bi{r})=w^A_0(\bi{r}-\bi{R}^A_i)$ located at the site
$i$ in sublattice $A$, and  $b_j  $  is    the annihilation operator
destructing  a particle in the Wannier state
$w^B_j(\bi{r})=w^B_0(\bi{r}-\bi{R}^B_j)$ located at the site $j$ in
sublattice $B$;
 the subscript $i\equiv (i_x,i_y)$ is the coordinate for the
lattice sites; $\hat{x}=(1,0)$ and
$\hat{y}=(0,1)$ represent the unit vectors in the $x$ and $y$ directions, respectively;  $t_x$ and $t_y$ are the amplitudes of hopping
 along the $x$ and $y$ directions, respectively. Without
loss of generality, we assume that $t_x$ and $t_y$ are positive in
the following process.

Based on the original model, we will consider two modified models
with different adding terms, respectively. In the first modified
model, we   add  a staggered
  potential to the original model. The   total
Hamiltonian can be described by $H_1=H_0+H_s$, where $H_s$ is the
staggered potential Hamiltonian as
\begin{eqnarray}
H_s=-v \sum_{i\in A} a^\dag_ia_i+v\sum_{j\in
B}b^\dag_jb_j,\label{H2}
\end{eqnarray}
where $v$ is the magnitude of the staggered  potential. In the second modified
model, we add   the diagonal hopping terms to the original model, which is shown in
Fig.\ref{fig1}(b).
 The Hamiltonian is $H_2=H_0+H_d$, where $H_d$ is the diagonal
 hopping Hamiltonian as
\begin{eqnarray}
H_d&=&-t_{xy}\sum_{i\in A} [ a^\dag_{i}  b_{i+\hat{x}+\hat{y}} +
a^\dag_{i}  b_{i-\hat{x}-\hat{y}} \nonumber\\&& -a^\dag_{i}
b_{i-\hat{x}+\hat{y}}  - a^\dag_{i}  b_{i+\hat{x}-\hat{y}}+ {\rm
H.c.}],
\end{eqnarray}
where $t_{xy}$ is the amplitude of hopping  along the diagonal
direction.

For the original model and the two modified models, the lattice vectors can be expressed as $\bi{R}_n=n_1
\bi{a}_1+n_2\bi{a}_2$ with $n_1$ and $n_2$ being integers. We chose
a lattice site in sublattice $A$ as the origin, then the lattice
sites in sublattice $A$ and sublattice $B$ can be written as
$\bi{R}^A_n=\bi{R}_n$ and $\bi{R}^B_n=\bi{R}_n+\bi{a}_1/2$.
 In the reciprocal
lattice, the reciprocal lattice vectors are defined as
$\bi{K}_m=m_1\bi{b}_1 +m_2\bi{b}_2$, where $m_1$ and $m_2$ are
integers, and $\bi{b}_1= (\pi,0)$ and $\bi{b}_2=(0,2\pi)$ are the
corresponding primitive reciprocal-lattice vectors. Thus, the
Brillouin zone is $-\pi/2 \leq k_x\leq \pi/2, -\pi \leq k_y\leq\pi$,
as shown in Fig.\ref{fig2}(b).

\section{Massless Dirac fermions, Moving and merging of Dirac points}
 In this section, we show that Dirac points exist in the original model. In the modified model with a staggered potential, the Dirac points in
 the Brillouin zone move when the magnitude of the staggered potential changes,  and they merge when the magnitude of the staggered potential arrives at a critical value.
 In the modified model with the diagonal hopping terms, the two Dirac points in the Brillouin zone move towards two opposite directions as the amplitude
 of the diagonal hopping increases, and they never vanish or merge for any value of the amplitude of the diagonal hopping.
\subsection{The original model}

\begin{figure}[ht]
\includegraphics[width=0.55\columnwidth]{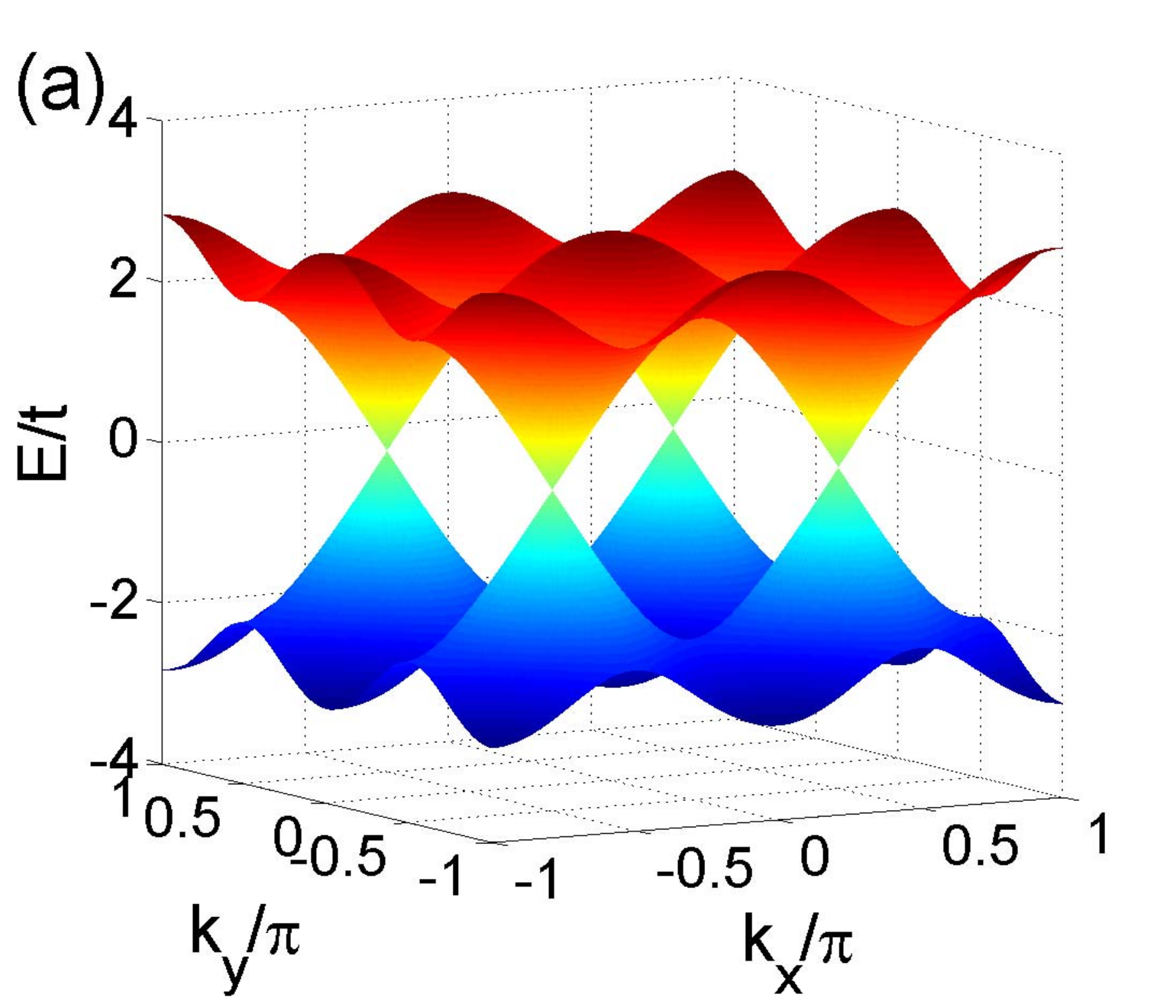}
\includegraphics[width=0.36\columnwidth]{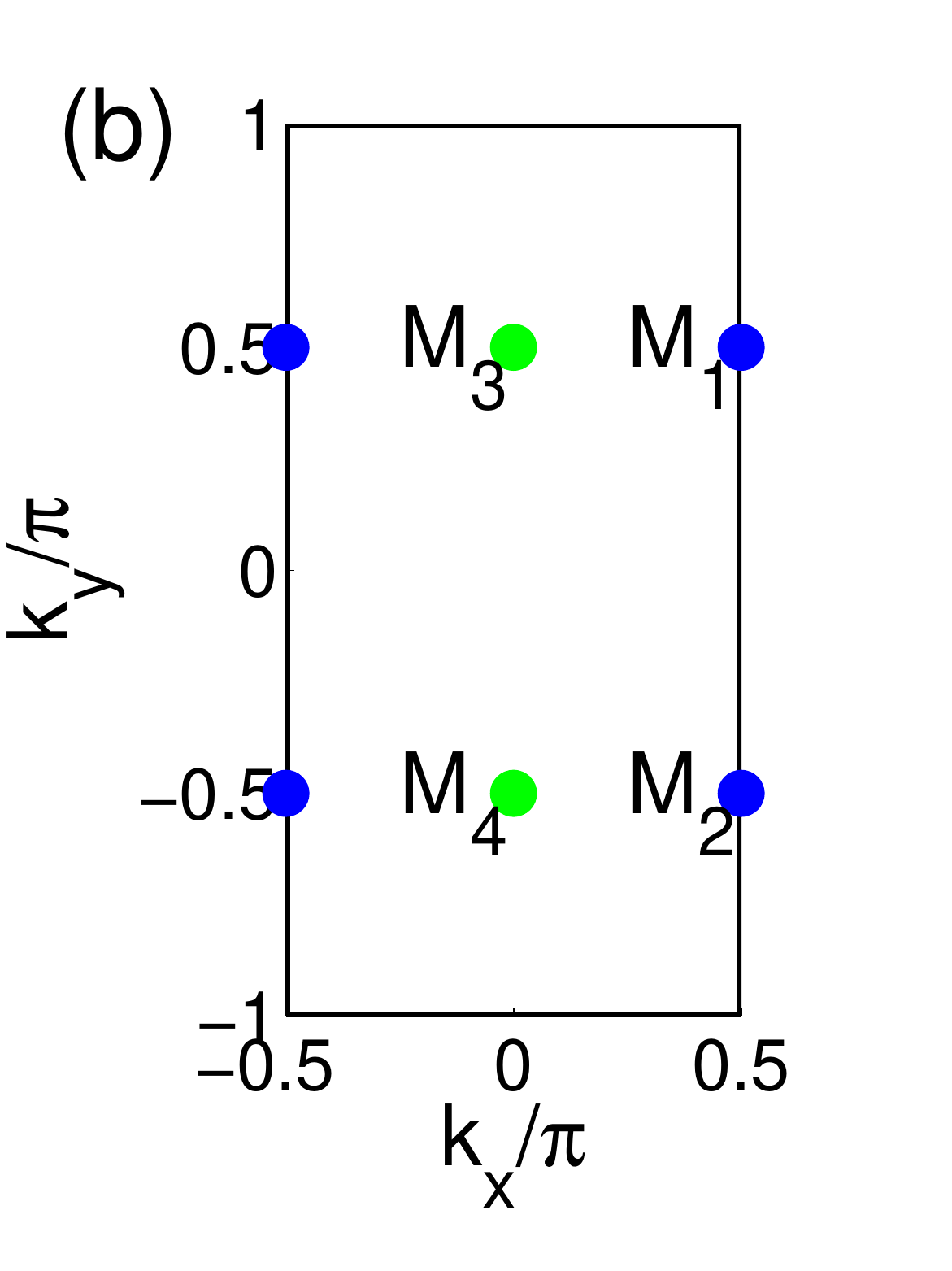}
\caption{(Color online). (a) The dispersion relation and (b) the
Brillouin zone for   the original model. Here, the
 filled circles represent the $\Upsilon$-invariant
points $M_{1}, M_2, M_3$ and $M_4$. The band
degeneracy occurs at $M_{1,2}$ marked by the blue color and
does not  at $M_{3,4}$ marked by the green color.} \label{fig2}
\end{figure}

  For the original model, we take the Fourier's transformation to the annihilation operators as
\begin{eqnarray}
{a}_{\bi{k}}&=&\frac{1}{\sqrt{N}}\sum_{i}
  {a}_{i}e^{-i\bi{k}\cdot\bi{R}^A_i},\label{fta}\\
  {b}_{\bi{k}}&=&\frac{1}{\sqrt{N}}\sum_{j }
  {b}_{j}e^{-i\bi{k}\cdot\bi{R}^B_j},\label{ftb}
\end{eqnarray}
   and define the two-component annihilation operator as $
 {\eta}_{\bi{k}}\equiv  [
 {a}_{\bi{k}},
 {b}_{\bi{k}} ]^T$.
The total Hamiltonian $H=H_0+H_1$ can be rewritten as $
H=\sum_{\bi{k}} {\eta}_{\bi{k}}^\dag{\cal H}_0(\bi{k})
{\eta}_{\bi{k}} $ with
\begin{eqnarray}
{\cal H}_0(\bi{k}) &=&-2t_x\cos k_x \sigma_x-2t_y\cos k_y\sigma_z,
\label{BH0}
\end{eqnarray}
and $\sigma_x$ and $\sigma_z$ are the Pauli matrices.

 Diagonalizing Eq.(\ref{BH0}), we obtain the dispersion relation as
\begin{eqnarray}
E_0(\bi{k})=\pm  \sqrt{4t_x^2\cos^2 k_x+4t_y^2\cos
k_y^2}.\label{dis}
\end{eqnarray}
 The corresponding  Bloch functions can be
expressed as
\begin{eqnarray}
\Psi^{(0)}_\bi{k}(\bi{r}) \equiv \left(\matrix{u^{(0)}_{1,\bi{k}}(\bi{r})\cr
u^{(0)}_{2,\bi{k}}(\bi{r})}\right)e^{i\bi{k}\cdot\bi{r}},\label{Bloch}
\end{eqnarray}
where $u^{(0)}_{i,\bi{k}}(\bi{r})=u^{(0)}_{i,\bi{k}}(\bi{r}+\bi{R}_n)$. In the momentum space, the
Bloch function $\Psi^{(0)}_\bi{k}(\bi{r})$ and eigenenergy $E_0(\bi{k})$ are
periodic for reciprocal lattice vectors, i.e.
$\Psi^{(0)}_\bi{k}(\bi{r})=\Psi^{(0)}_{\bi{k}+\bi{K}_m}(\bi{r})$ and
$E_0(\bi{k})=E_0(\bi{k}+\bi{K}_m)$.

The conduction and valence bands touch at
$(\pm \pi/2, \pm\pi/2)$, which are located at the boundary of the
Brillouin zone as shown in Fig.\ref{fig2}(a). Among these degenerate
points, there are only two distinct ones.  Near these degenerate
points, the single-particle Hamiltonian (\ref{BH0}) can be linearized
as
\begin{eqnarray}{h}(\bi{p})&=&  2t_x
 p_x\sigma_x\pm 2t_y    p_y\sigma_z, \label{lh}
\end{eqnarray}
where the signs $\pm$ representing the linearized Hamiltonian around
the different touching points,   respectively. Around the touching
points, the quasiparticles behave like  massless Dirac fermions. For
these massless Dirac fermions, a chirality can be defined as
\cite{Hou1}
\begin{eqnarray}
 w=\textrm{sgn}[\det(v_{ij})]=\pm 1,
 \label{ch}
 \end{eqnarray} for
a two-dimensional Hamiltonian
$h(\bi{k})=\sum_{ij}v_{ij}k_i\sigma_j$, with $\bi{k}$ and
$\bi{\sigma}$ being the  wave vector and the Pauli matrix in two dimensions,
respectively.  If we use   $\sigma_y$  to redenote
  $\sigma_z$ in Eq.({\ref{lh}), the
corresponding quasiparticles have a chirality $\pm 1$ as defined
above.    The quasiparticles are massless Dirac fermions with a
chirality, so they   can be considered as two-dimensional Weyl
fermions.   The chirality of  Dirac points can be considered as a
topological charge.

\begin{figure}[ht]
\includegraphics[width=0.45\columnwidth]{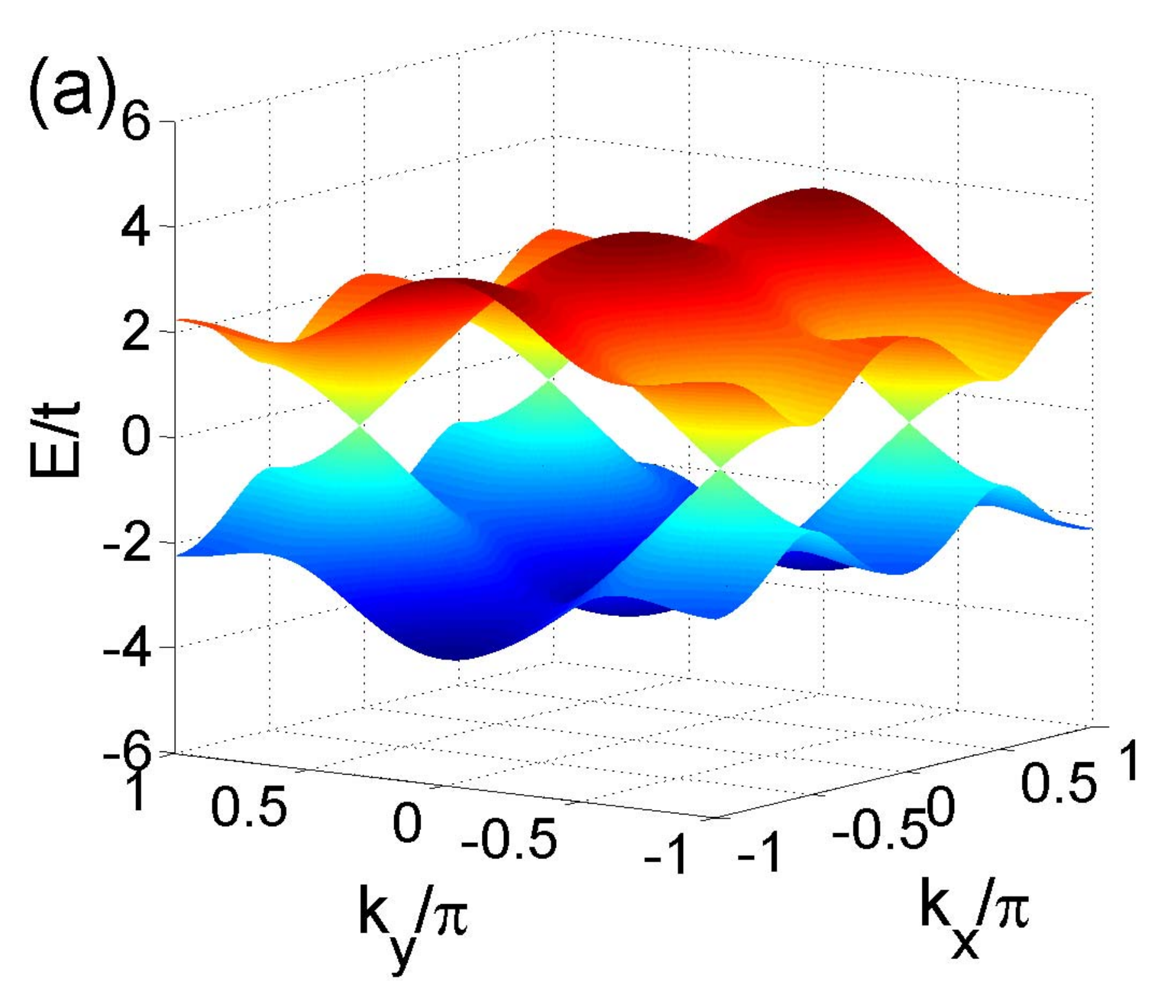}
\includegraphics[width=0.45\columnwidth]{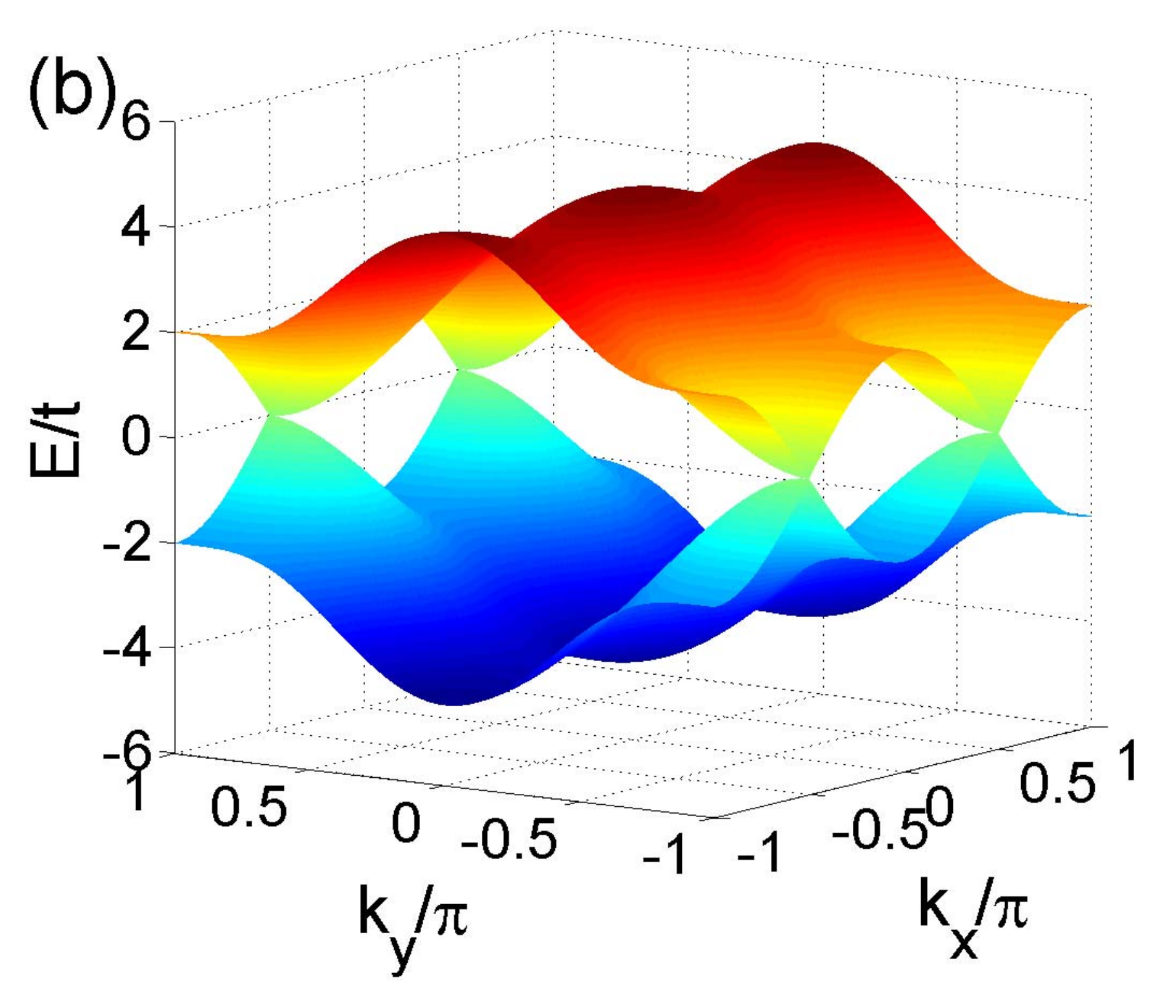}
\includegraphics[width=0.45\columnwidth]{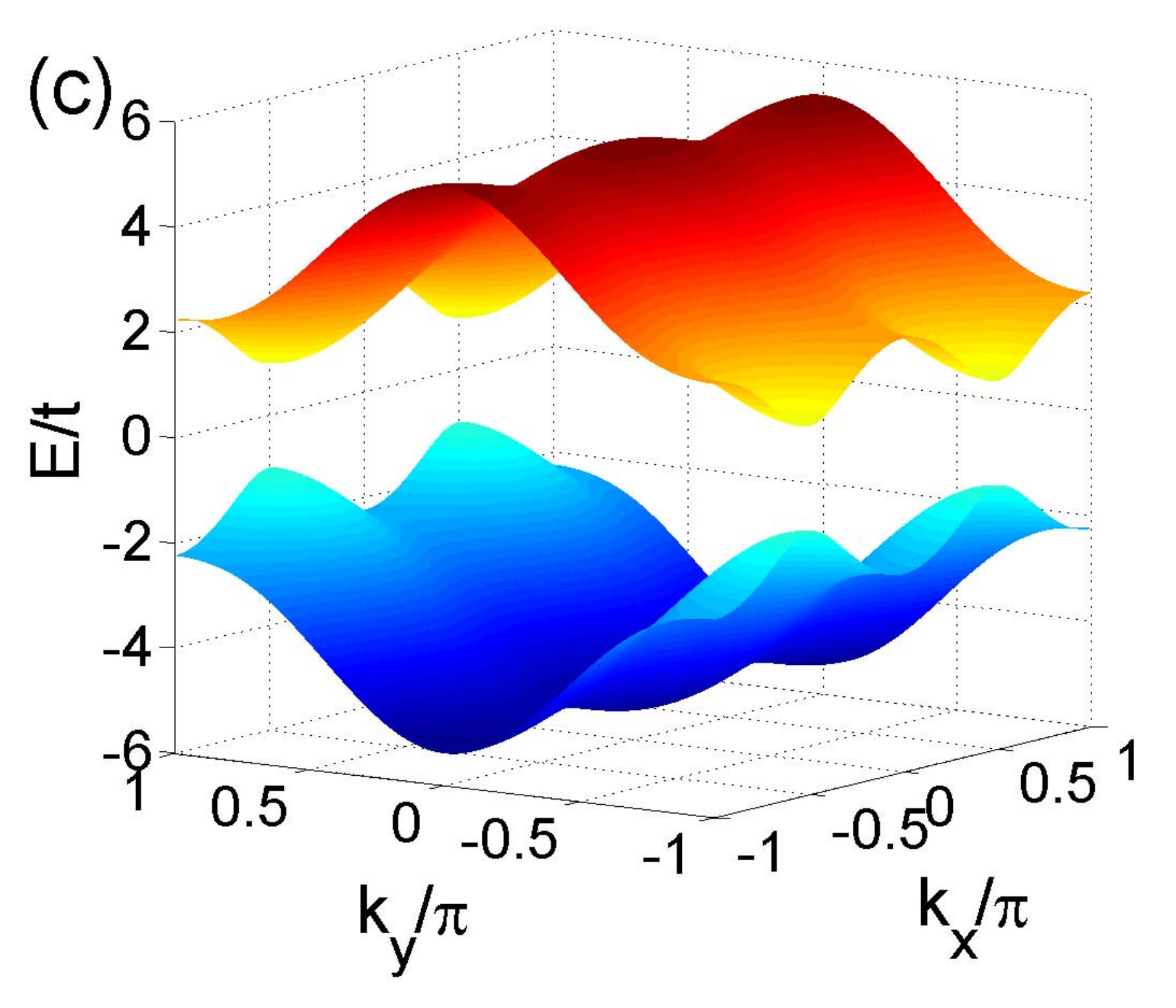}
\includegraphics[width=0.45\columnwidth]{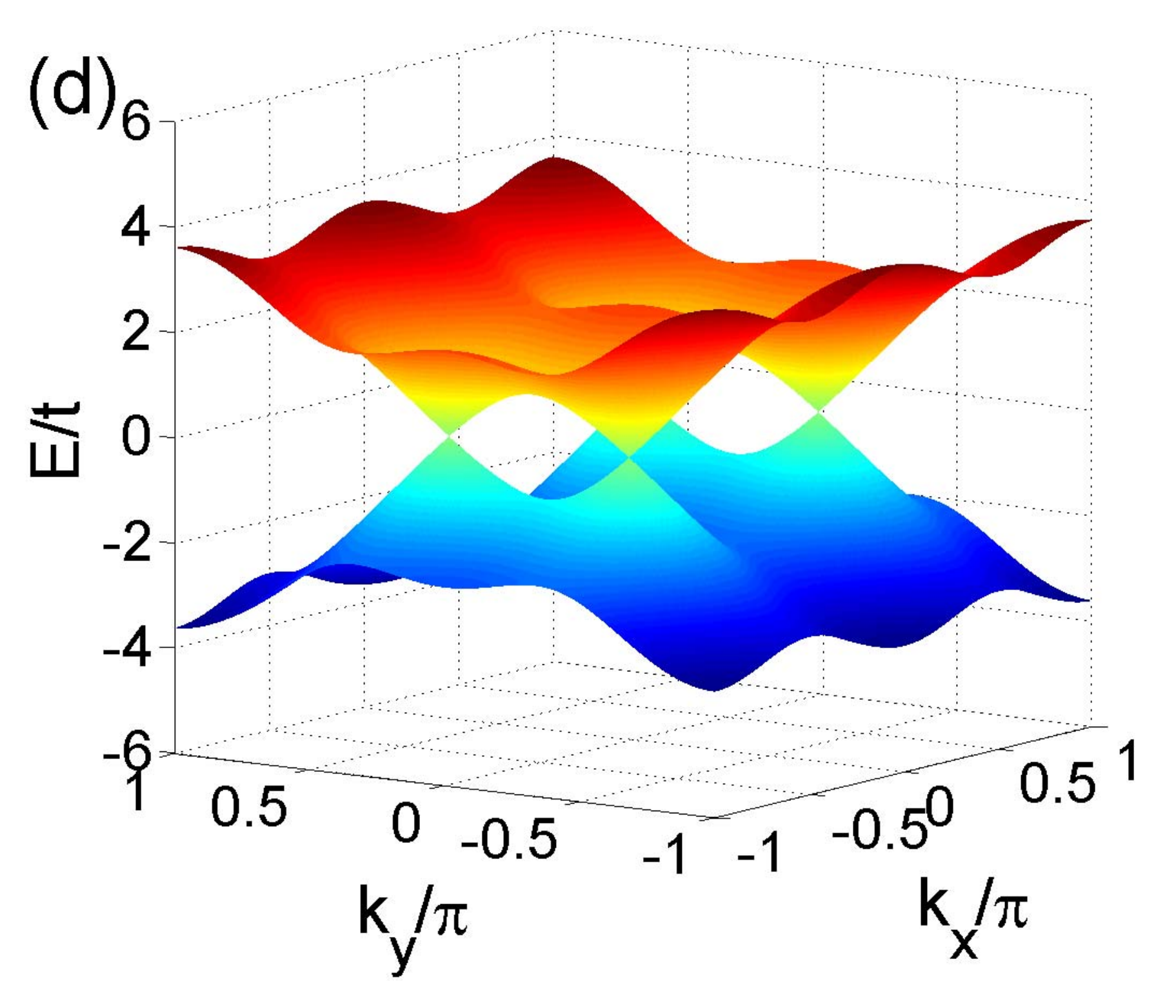}
\includegraphics[width=0.45\columnwidth]{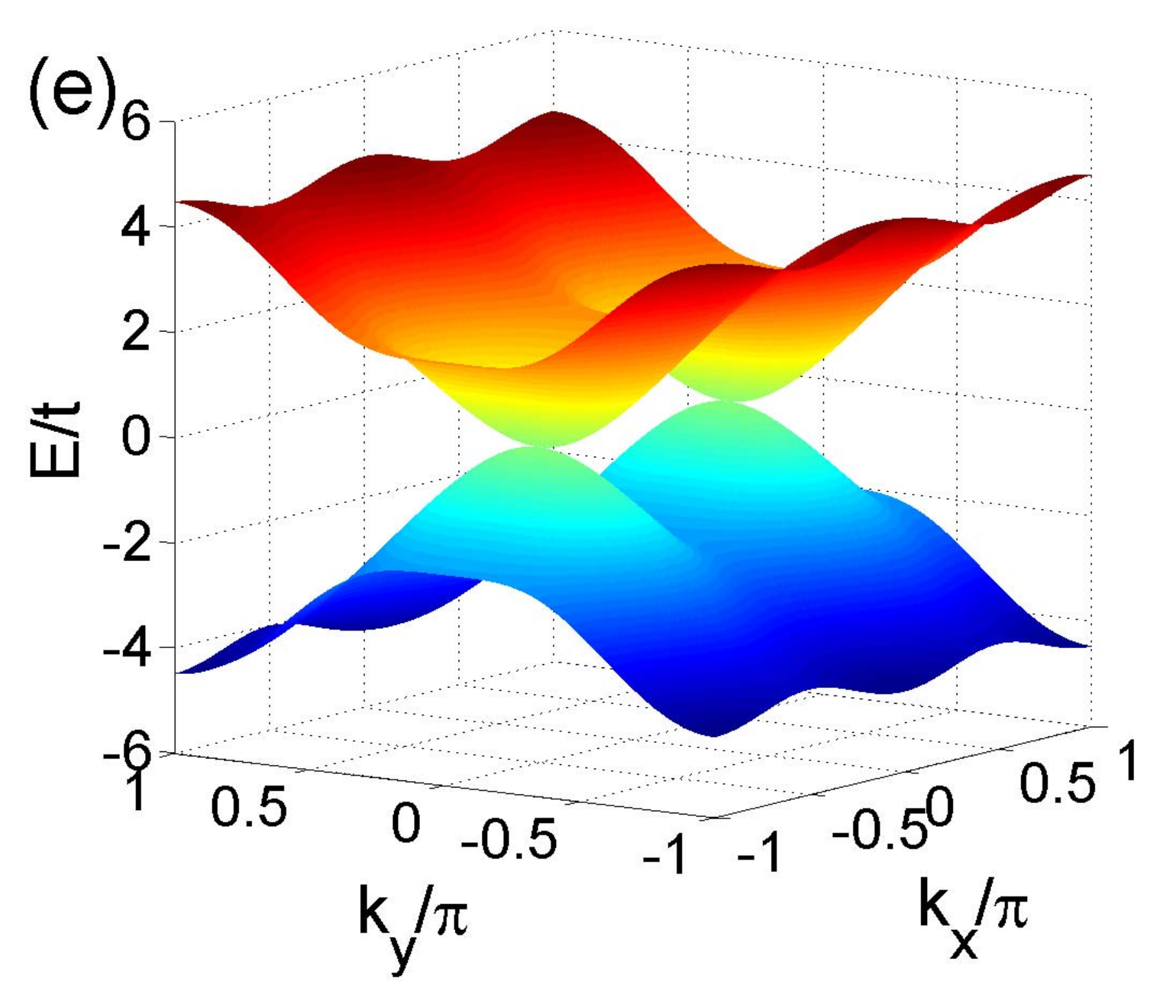}
\includegraphics[width=0.45\columnwidth]{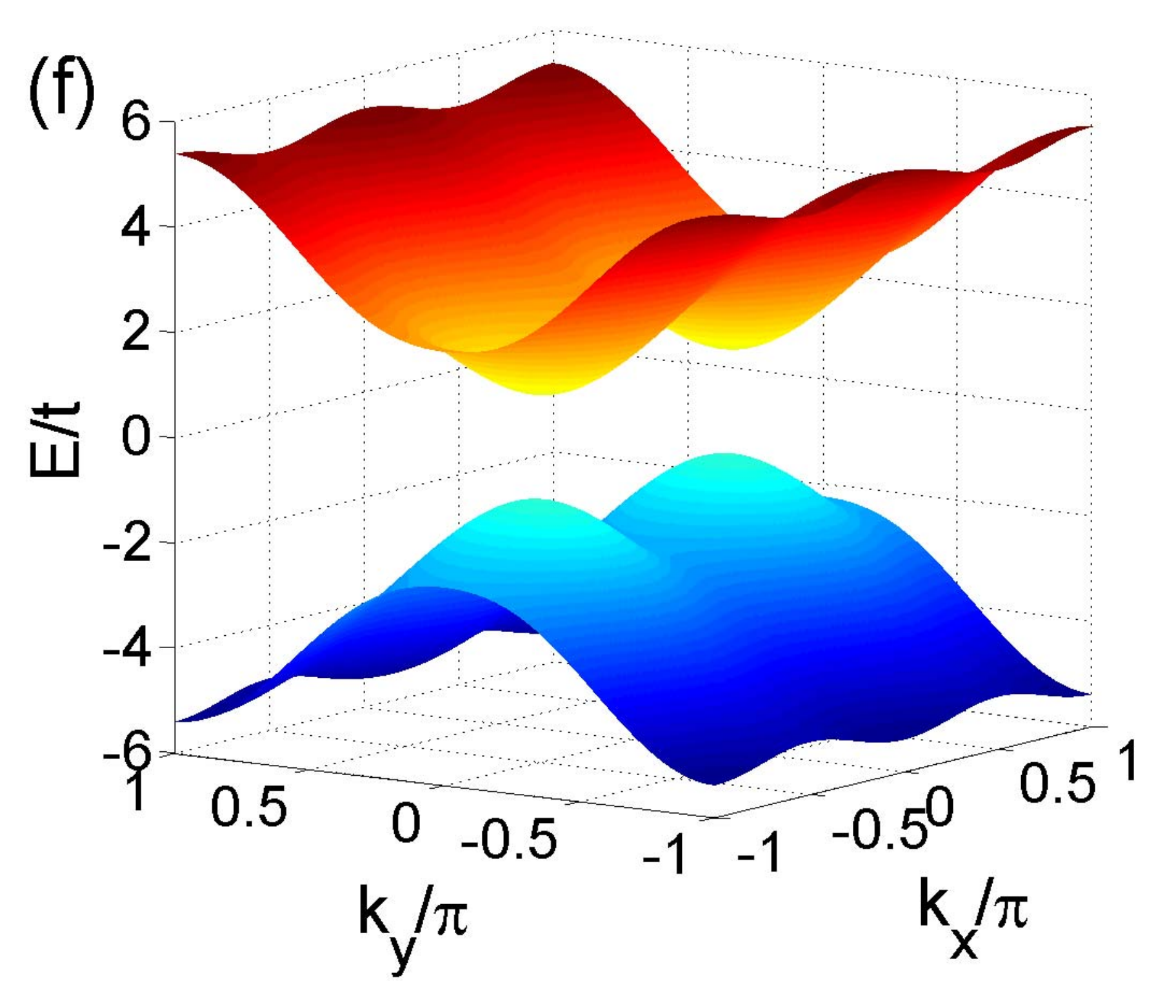}
\caption{(Color online). The dispersion relation  of  the modified
model with a staggered potential with $t_x=t_y=t$ and  (a) $v=t_y$,
(b) $v=2t_y$, (c) $v=3t_y$,(d) $v=-t_y$, (e) $v=-2t_y$,  (f)
$v=-3t_y$ } \label{fig3}
\end{figure}

\subsection{The modified model with a staggered potential}
For the modified model with a staggered potential, the Bloch
Hamiltonian can be written as
\begin{eqnarray}
{\cal H}_1(\bi{k}) &=&-2t_x\cos k_x \sigma_x-(v+2t_y\cos k_y)
\sigma_z,\label{BH1}
\end{eqnarray}
and the corresponding dispersion relation is
\begin{eqnarray}
E_1(\bi{k})=\pm\sqrt{4t_x^2\cos^2 k_x+(v+2t_y\cos
k_y)^2}.\label{dis1}
\end{eqnarray}
 In this model, the energy dispersion relation
(\ref{dis1})  possesses Dirac points at $(\pi/2,\pm
\arccos(-v/2t_y))$ in the Brillouin zone for $|v|<2t_y$,  as shown
Fig.(\ref{fig3})(a) and (d).
 We find that the two distinct Dirac points move  as  $v$ changes. For
the positive $v$, with   increasing   $v$, the two Dirac points move
away from each other.   When $v=0$, the Dirac points are located at
points $(\pi/2,\pm \pi/2)$ as shown in Fig.\ref{fig2}(a). When $v$
changes from $0$ to $t_y$, the two distinct Dirac points move to
$(\pi/2, \pm 2\pi/3)$, respectively, as shown in \ref{fig3}(a). When
$v$ arrives at $2t_y$, the two Dirac points move to  $(\pi/2,
\pm\pi)$, which are the same point    on the boundary of the
Brillouin zone as shown in Fig.\ref{fig3}(b), that is to say, the
Dirac points merge. If one continues to increase $v$ to $v>2t_y$, a
gap opens, as shown in Fig.\ref{fig3} (c). Then the system turns
into an insulator. For the negative $v$, with increasing   $|v|$,
the two Dirac points move towards each other. When $v$ changes from
$0$ to $-t_y$, the two distinct Dirac points move to $(\pi/2, \pm
\pi/3)$, respectively, as shown in \ref{fig3}(d). When $v$ arrives
at $-2t_y$, the two Dirac points merge at the point $(\pi/2, 0)$ as
shown in \ref{fig3}(e). When $v$ is less than $-2t_y$, a gap opens
and the systems turns into an insulator, as shown in
Fig.\ref{fig3}(f).

The above  merging process  of Dirac points can be interpreted from
the topological view. Two Dirac points have opposite chirality, that
is to say, they have opposite topological charges. As long as the
band touching points are protected by a symmetry, the topological
charges can not be destroyed, and the system is a topological
semimetal. However, when the Dirac points with opposite topological
charges meet, they merge and the opposite topological charges
annihilate each other.\cite{Volovik}  A further increase of $|v|$
makes a gap open and the system turns into an insulator.

\subsection{The modified model with the diagonal hopping terms}
\begin{figure}[ht]
\includegraphics[width=0.45\columnwidth]{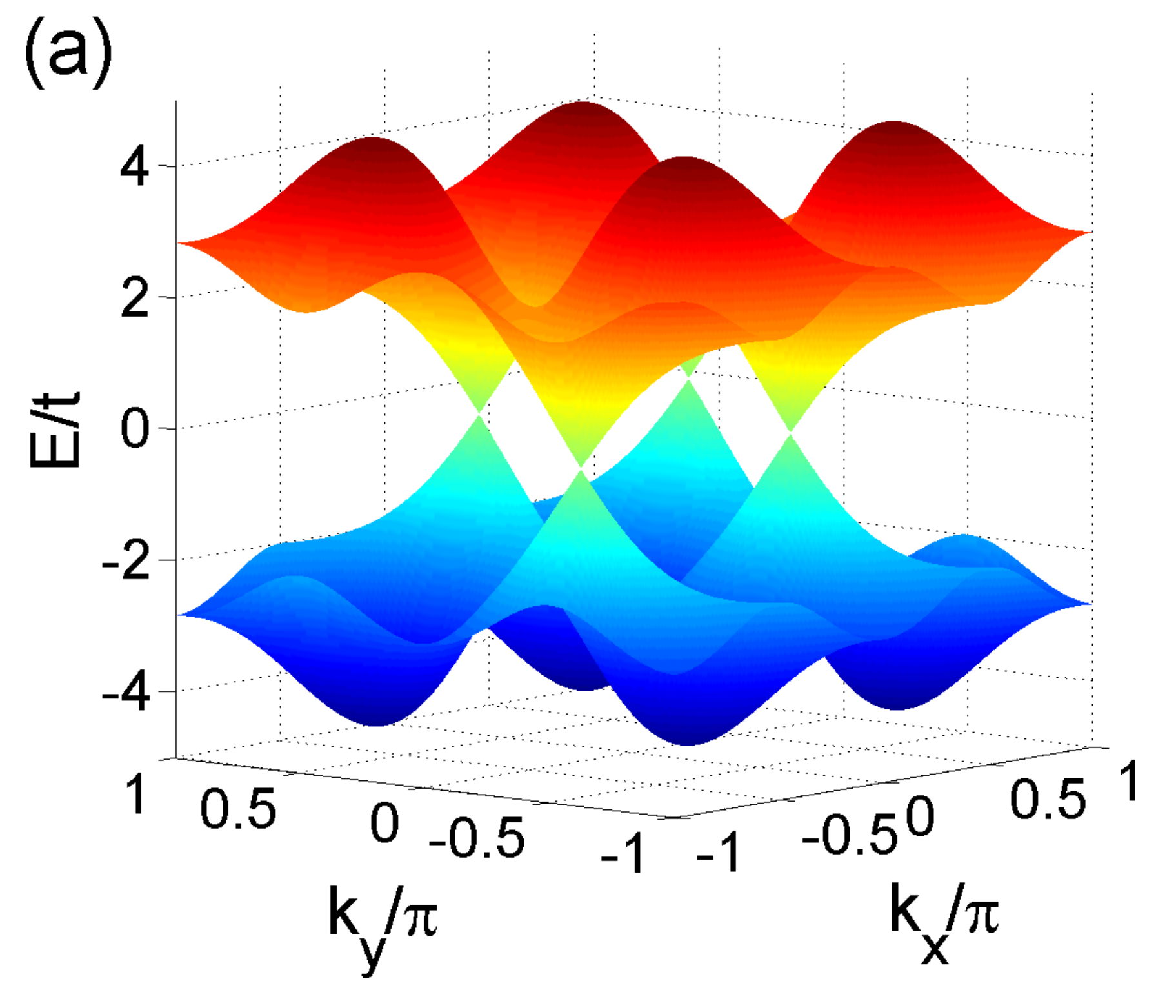}
\includegraphics[width=0.45\columnwidth]{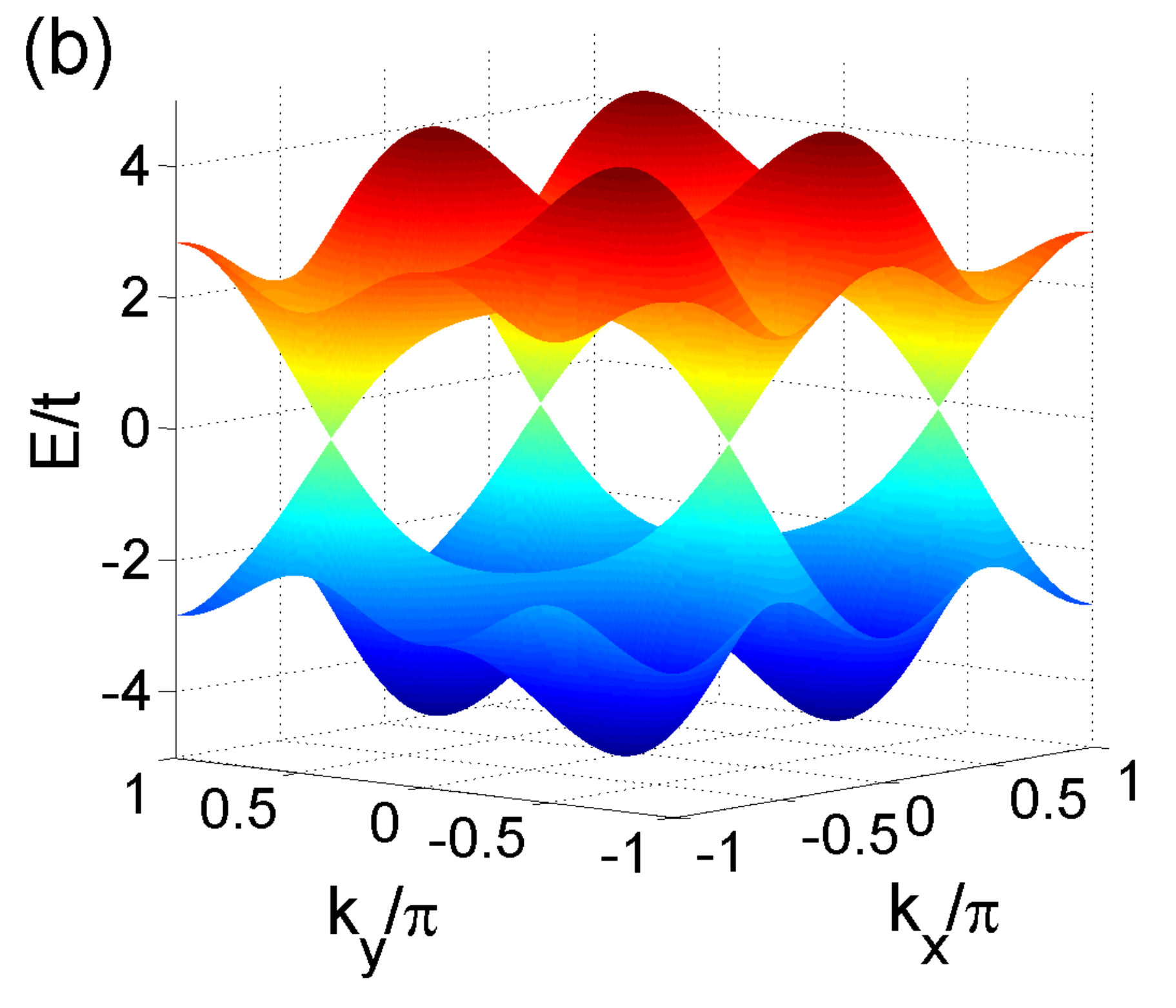}
\caption{(color online). The dispersion relation  of  the modified
model with the diagonal hopping terms with $t_x=t_y=t$ and (a)
$t_{xy}=t_x$, (b) $t_{xy}=-t_x$. } \label{fig4}
\end{figure}
For the modified model with the diagonal hopping terms, the Bloch
Hamiltonian can be expressed as
\begin{eqnarray}
{\cal H}_2(\bi{k})&=&-(2t_x\cos k_x+4t_{xy}\sin k_x \sin k_y)
\sigma_x\nonumber\\
&&-2t_y\cos k_y\sigma_z, \label{BH2}
\end{eqnarray}
and the corresponding dispersion relation is
\begin{eqnarray}
E_2(\bi{k})=\pm  \sqrt{(2t_x\cos k_x+4t_{xy}\sin k_x\sin
k_y)^2+4t_y^2\cos^2k_y}.\nonumber\\
\end{eqnarray}
In this model, the bands touch at the points
$(-\arctan(t_x/2t_{xy}),\pi/2)$ and $(\arctan(t_x/2t_{xy}), -\pi/2)$
in the Brillouin zone. Near these degenerate points, the dispersion
relation is linear. Thus, these points are also Dirac points and
have a chirality defined above. Compared with the original model,
the Dirac points have a shift in the Brillouin zone. For the case
$t_{xy}/t_x>0$, with increasing   $|t_{xy}|$, the Dirac point in the
$k_y>0$ half part of the Brillouin zone moves towards the $x$
direction, while the Dirac point in the $k_y<0$ half part of the
Brillouin zone moves towards  the negative $x$ direction as shown in
Fig.\ref{fig4}(a). When the parameter $|t_{xy}|$ approaches
infinity, the Dirac points move asymptotically  to the $k_x=0$ line.
For the case of $t_{xy}/t_x<0$, the Dirac points have similar shifts
but with the opposite directions compared with the case of
$t_{xy}/t_x>0$. That is to say, the Dirac point in the $k_y>0$ half
part of the Brillouin zone moves towards the negative  $x$
direction, while the Dirac point in the $k_y<0$ half part of the
Brillouin towards the  $x$ direction   as shown in
Fig.\ref{fig4}(b).  For any value of the parameter $t_{xy}$, the
system remains gapless and the Dirac points never merge, which is
different from the modified model with a staggered potential.

\section{Explanation from  hidden symmetry protection}

In this section, we prove that the Dirac points   are protected by a
kind of hidden symmetry in the original model.  In  the two modified
models, the additive terms violate the hidden symmetry respected by
the original model. Thus, we   develop a mapping method to find
hidden symmetries evolving with the parameters for the two modified
model. We  explain the moving  of Dirac points in the two modified
models by the evolution of the hidden symmetries along with the
variation of the parameters,  and we explain the merging of Dirac
points in the modified model with a staggered potential by the
disappearance of the hidden-symmetry-invariant points in the
Brillouin zone.

\subsection{The original model }

The original model supports the existence of massless Dirac fermions
with the Dirac points located at $(\pi/2, \pm\pi/2)$ in the
Brillouin zone. We will show that  the band degeneracies at the
Dirac points are protected by a hidden symmetry.
  Here, we define a hidden symmetry with the operator
  as follows,
\begin{eqnarray}
\Upsilon=(e^{i\pi})^{i_y}\sigma_x KT_{{\bi{a}}_1/2}, \label{sym}
\end{eqnarray}
where $T_{\bi{a}_1/2}=\{E|\bi{a}_1/2\}$ is a translation operator
that moves the lattice by $\bi{a}_1/2$  along the $x$ direction, $K$
is the complex conjugate operator, $\sigma_x$ is the Pauli matrix
representing sublattice exchange, and $(e^{i\pi})^{i_y}$ is a local
$U(1)$ gauge transformation. Obviously, the hidden symmetry operator
$\Upsilon$ is an antiunitary operator.  The corresponding inverse
operator is $ \Upsilon^{-1}=(e^{i\pi})^{i_y}\sigma_x
KT_{{\bi{a}}_1/2}^{-1}$. It is easy to verify that the Hamiltonian
of the original model is invariant under the hidden symmetry
transformation, i.e. $H_0=\Upsilon H_0\Upsilon^{-1}$.

The hidden symmetry operator $\Upsilon$ acts on the Bloch functions
(\ref{Bloch}) as follows
\begin{eqnarray}
\Upsilon\Psi^{(0)}_{\bi{k}}(\bi{r})&=&\left(\matrix{u^{(0)*}_{2,\bi{k}}(\bi{r}-\bi{a}_1/2)e^{ik_x}\cr
u^{(0)*}_{1,\bi{k}}(\bi{r}-\bi{a}_1/2)e^{ik_x}}\right)e^{-i[k_xx+(k_y-\pi)y]}\nonumber\\
&=&{\Psi^{(0)}_{\bi{k}'}}'(\bi{r}).\label{Bloch2}
\end{eqnarray}
 Because $\Upsilon$  is the symmetry operator for the original model,
${\Psi^{(0)}_{\bi{k}'}}'(\bi{r})$ must be a Bloch wave function of
the original model. Thus, we obtain
$u^{(0)}_{1,\bi{k}'}(\bi{r})=u^{(0)*}_{2,\bi{k}}(\bi{r}-\bi{a}_1/2)e^{ik_x}$
and
$u^{(0)}_{2,\bi{k}'}(\bi{r})=u^{(0)*}_{1,\bi{k}}(\bi{r}-\bi{a}_1/2)e^{ik_x}$,
$k'_x=-k_x$ and $k'_y=\pi-k_y$. The square of the hidden symmetry
operator is
\begin{eqnarray}
\Upsilon^2=T_{ {\bi{a}}_1},\label{A2}
\end{eqnarray}
where $T_{ \bi{a}_1}=\{E| \bi{a}_1\}$. Therefore, we have
\begin{eqnarray}
\Upsilon^2\Psi^{(0)}_{\bi{k}}(\bi{r})=T_{\bi{a}_1}\Psi^{(0)}_{\bi{k}}(\bi{r})=e^{-2ik_x}\Psi^{(0)}_{\bi{k}}(\bi{r}).
\label{U2}
\end{eqnarray}

From Eqs. (\ref{Bloch}) and (\ref{Bloch2}), it is easy to show that
the operator $\Upsilon$ has the following effect when acting on the
wave vector $\bi{k}$:
 \begin{eqnarray}
 \Upsilon: \bi{k}=(k_x,k_y)\rightarrow \bi{k}'=(-k_x, -k_y+\pi).
 \label{UT}
 \end{eqnarray}
If $\bi{k}'=\bi{k}+\bi{K}_m$, then we can say that $\bi{k}$ is an
 invariant point under the hidden symmetry
transformation. In the Brillouin zone, the $\Upsilon$-invariant
points are $M_{1,2}=( \pi/2, \pm \pi/2)$  and $M_{3,4}=(0,\pm\pi/2)$
as shown in Fig.\ref{fig2}(b). For a $\Upsilon$-invariant point
$M_i$,
 we have $
\Upsilon\Psi^{(0)}_{M_i}(\bi{r})={\Psi^{(0)}_{M_i}}'(\bi{r})$.  Thus,
$\Psi^{(0)}_{M_i}(\bi{r})$ and  ${\Psi^{(0)}_{M_i}}'(\bi{r})$ are both
the eigenstates of Hamiltonian $H_0$ and  have the same eigenenergy
$E_0(M_i)$. Considering Eq.(\ref{A2}), we have $
\Upsilon^2\Psi^{(0)}_{M_i}(\bi{r})=T_{\bi{a}_1}\Psi^{(0)}_{M_i}(\bi{r})=e^{-2iM_{ix}
}\Psi^{(0)}_{M_i}(\bi{r})$, where $M_{ix}$ is the $x$ component of
$M_i$. We define $(\psi, \varphi)$ as the inner product of the
two wave functions $\psi$ and $\varphi$. The antiunitary operator
$\Upsilon$ has the property that $(\Upsilon\psi,
\Upsilon\varphi)=(\psi, \varphi)^*=(\varphi, \psi)$. Therefore, we
have
\begin{eqnarray}
({\Psi^{(0)}_{M_i}}',\Psi^{(0)}_{M_i})&=&(\Upsilon\Psi^{(0)}_{M_i},\Upsilon{\Psi^{(0)}_{M_i}}')
=(\Upsilon\Psi^{(0)}_{M_i},\Upsilon^2\Psi^{(0)}_{M_i})\nonumber\\
&=&e^{-2iM_{ix}}( {\Psi^{(0)}_{M_i}}',
\Psi^{(0)}_{{M_i}}).\label{prot}
\end{eqnarray}
Substituting the concrete $\Upsilon$-invariant points $M_i$,   we
have $\Upsilon^2=-1$ at $M_{1,2}$, while $\Upsilon^2=-1$ at
$M_{3,4}$.  Then we obtain the solution
$({\Psi^{(0)}_{M_i}}',\Psi^{(0)}_{M_i})=0$ at  $M_{1,2}$, i.e.,
${\Psi^{(0)}_{M_i}}'$ and $\Psi^{(0)}_{M_i }$ are orthogonal to each
other, while $({\Psi^{(0)}_{M_i}}',\Psi^{(0)}_{M_i})$ is
unconstrained for Eq.(\ref{prot}) at $M_{3,4}$.  Therefore, we
arrive at the conclusion that the system must be degenerate at
points $M_{1,2}$
 in the
Brillouin zone, which are just the positions where the Dirac points
are located. We can conclude that the two Dirac points are protected
by the hidden symmetry $\Upsilon$.

\subsection{The modified model with a staggered  potential }

\begin{figure}[ht]
\includegraphics[width=0.325\columnwidth]{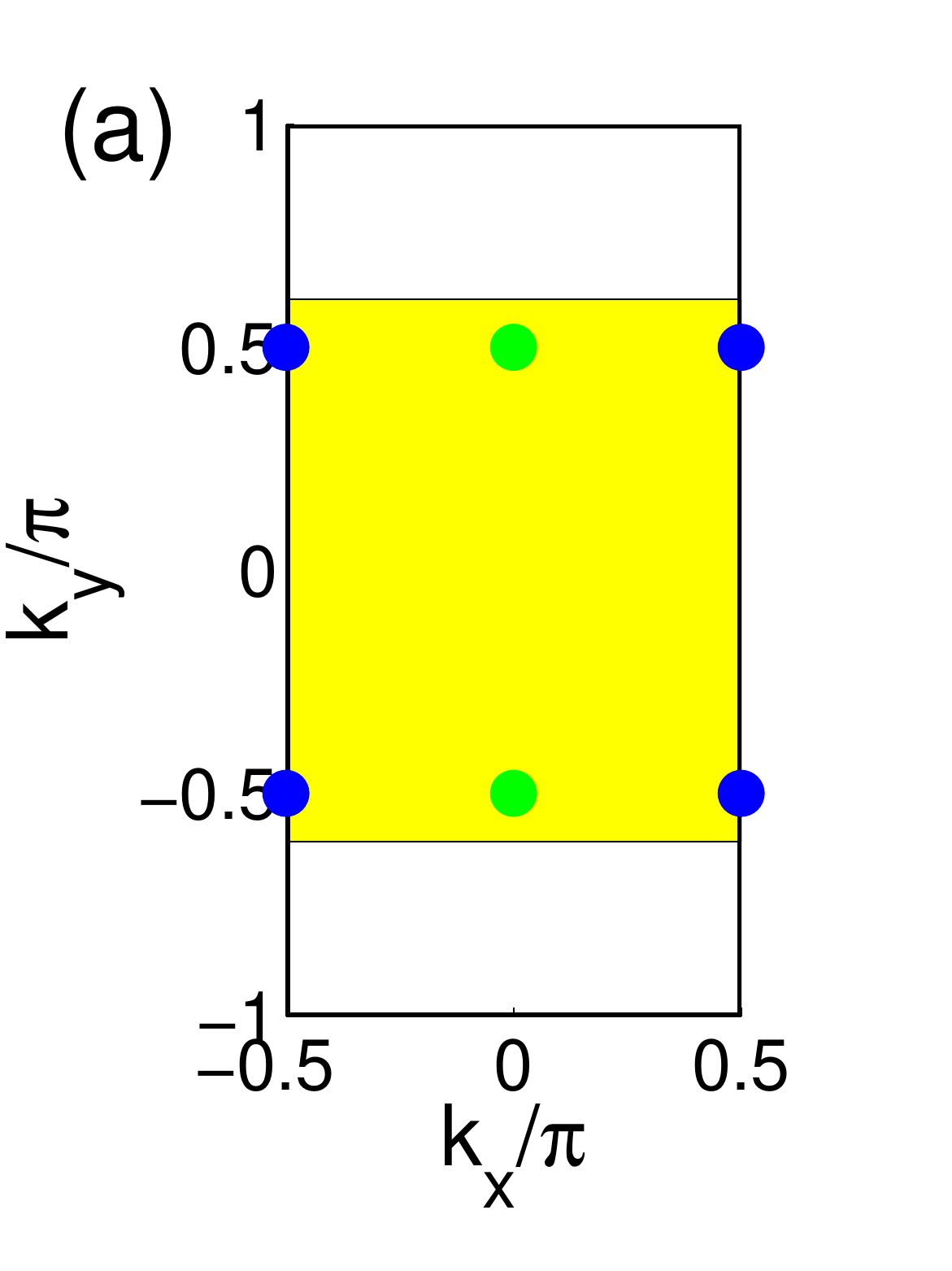}
\includegraphics[width=0.325\columnwidth]{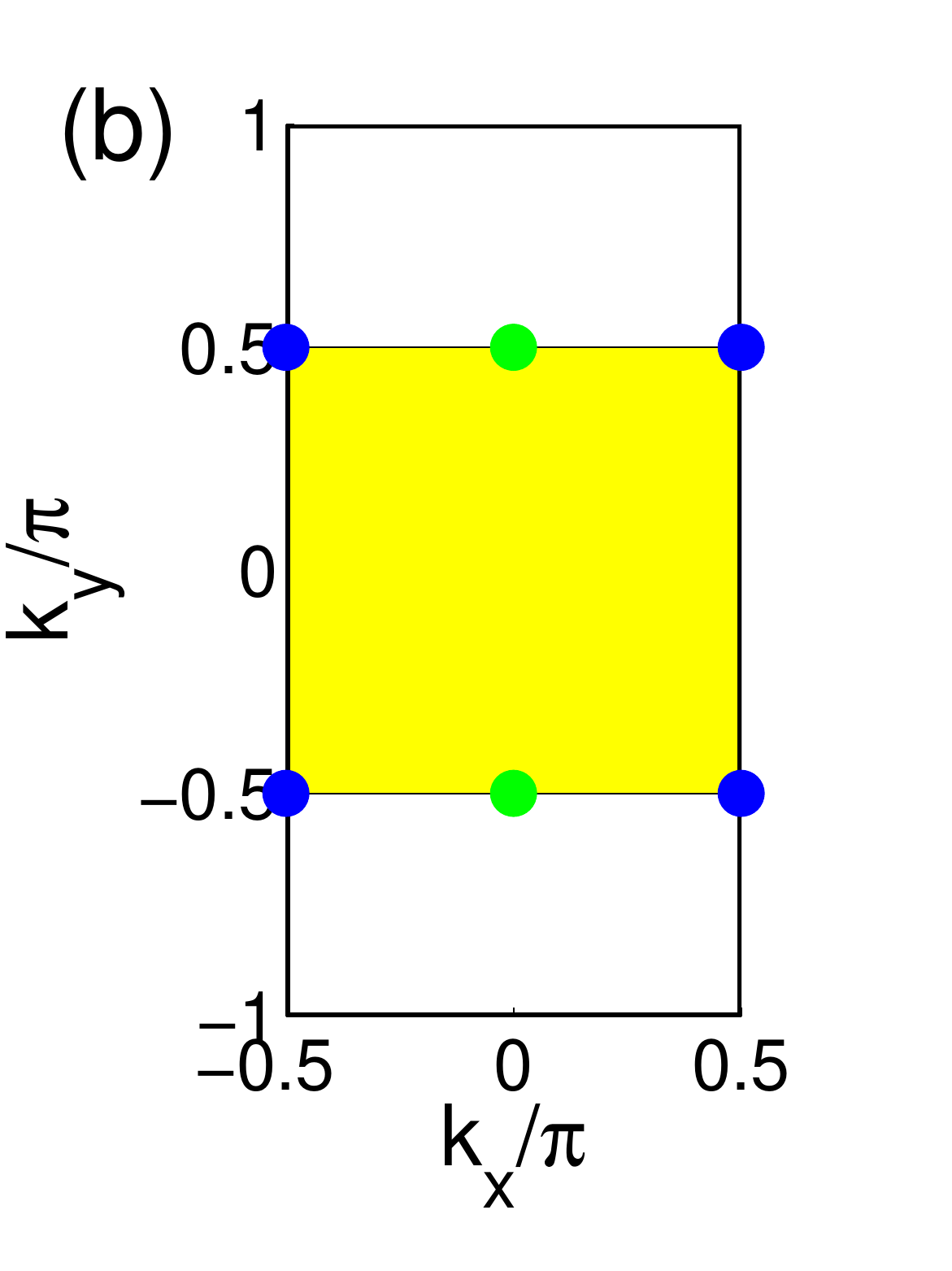}
\includegraphics[width=0.325\columnwidth]{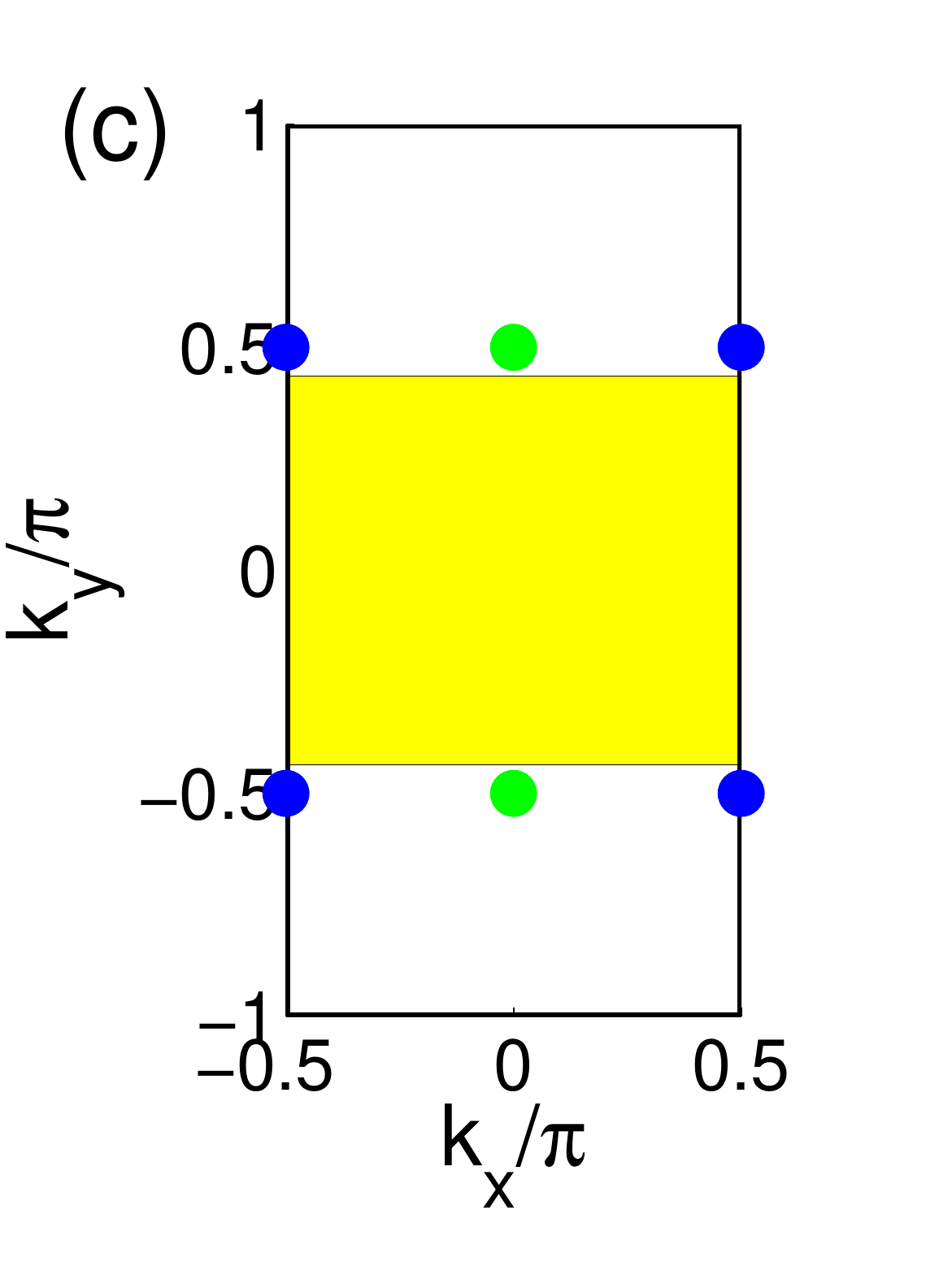}
\includegraphics[width=0.325\columnwidth]{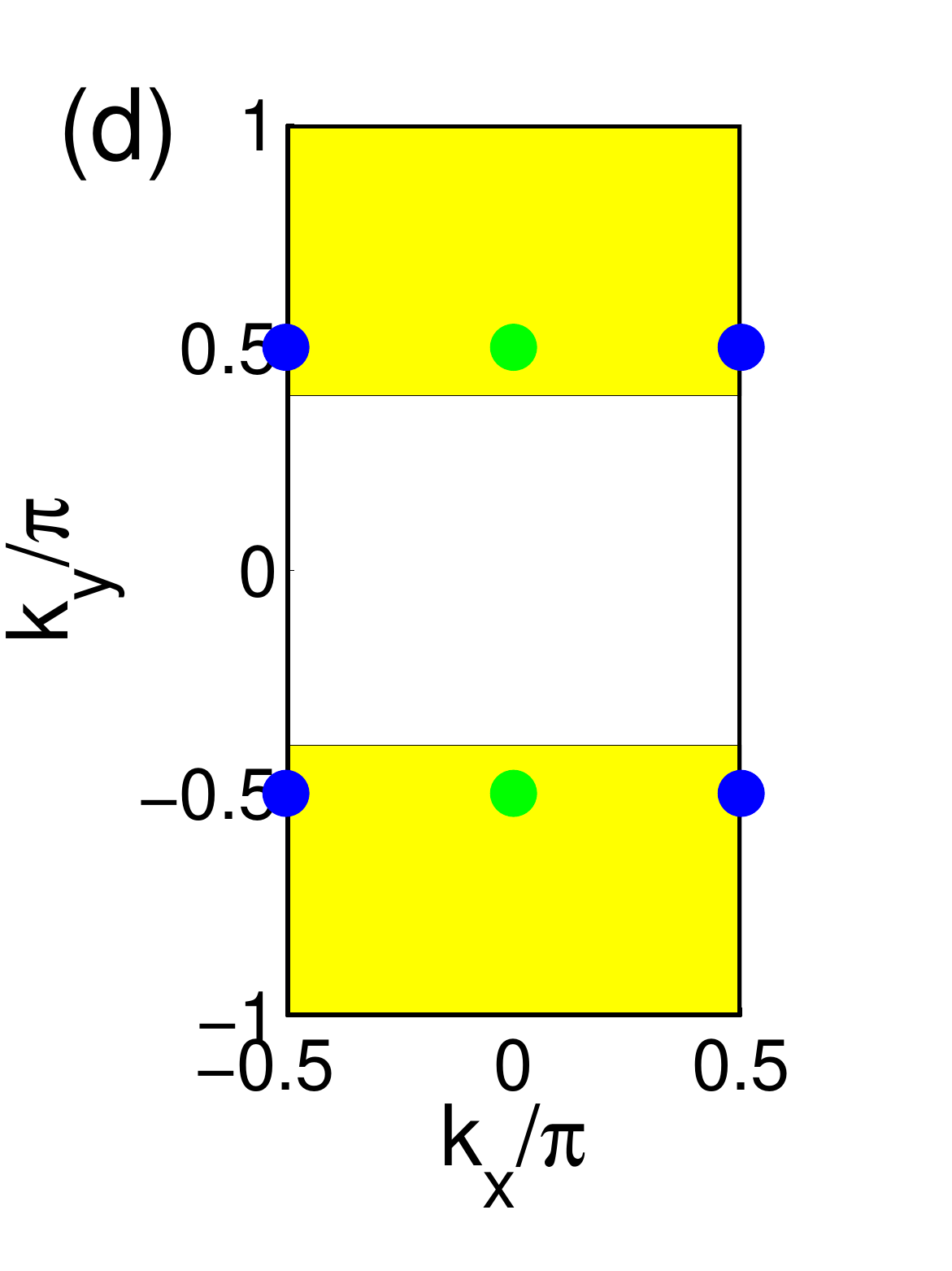}
\includegraphics[width=0.325\columnwidth]{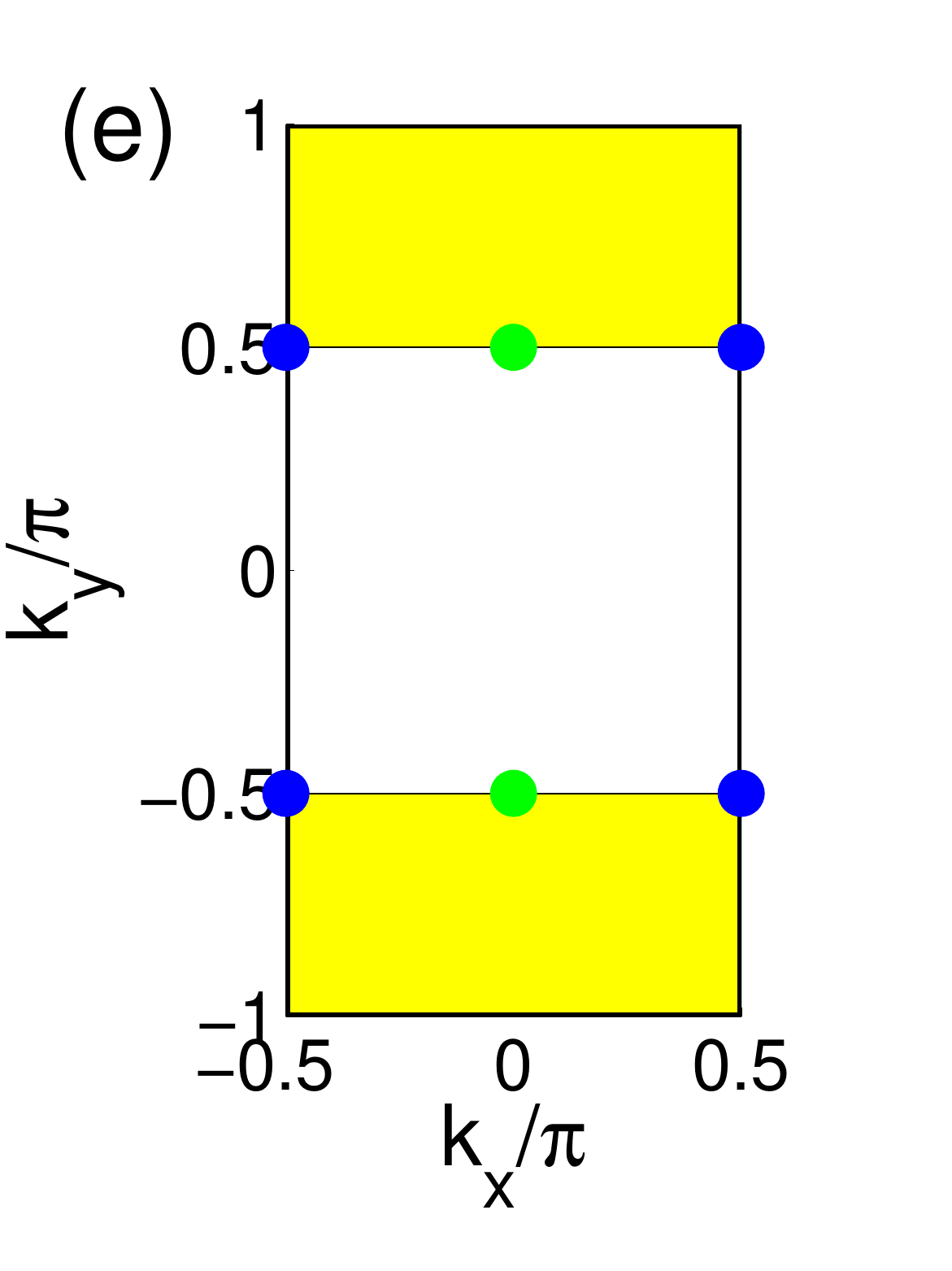}
\includegraphics[width=0.325\columnwidth]{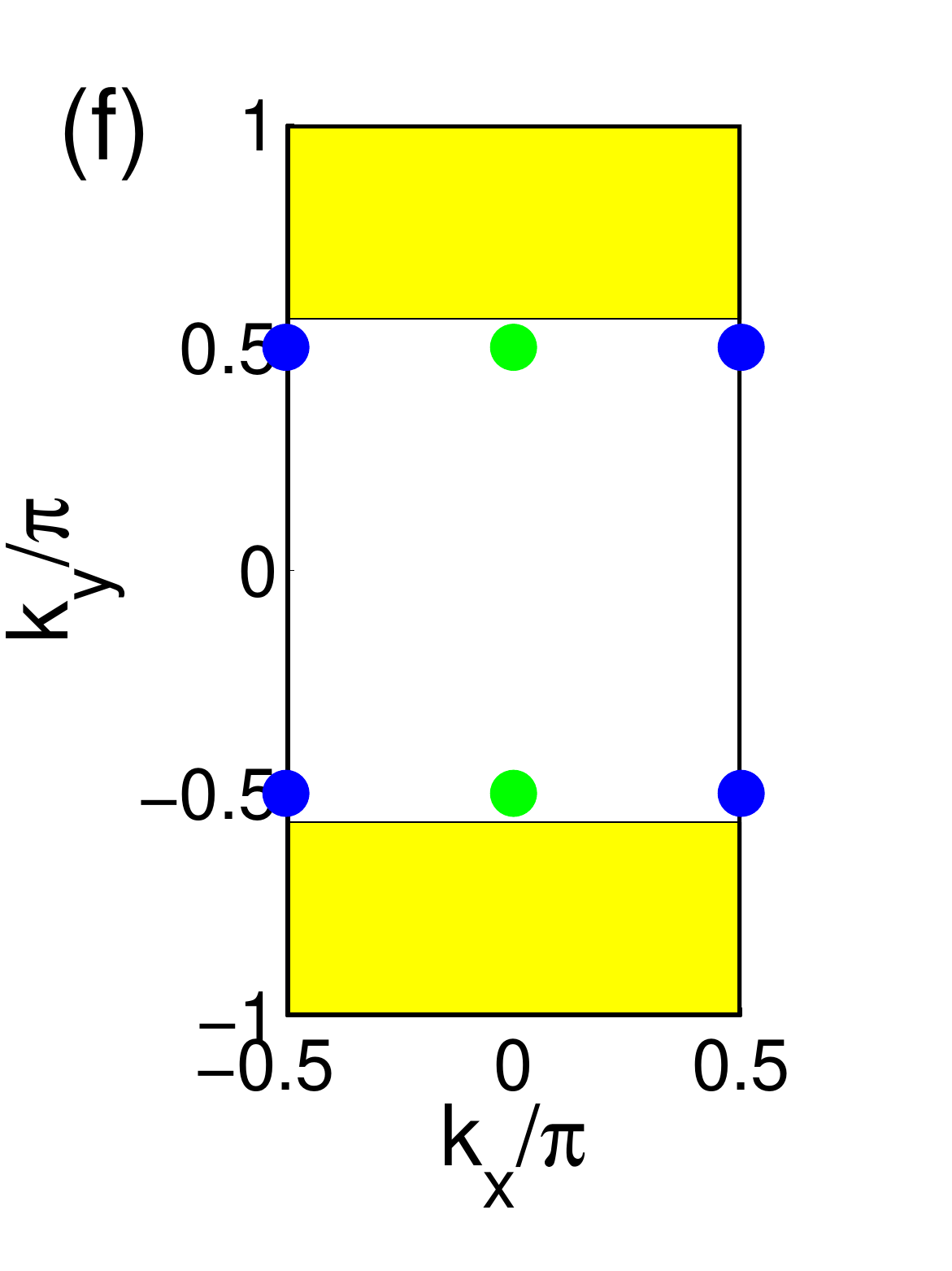}
\caption{(color online). The mapping of the Brillouin zone of the
modified model with a staggered potential   for (a) $v=t_y$, (b)
$v=2t_y$, (c) $v=3t_y$,(d) $v=-t_y$, (e) $v=-2t_y$, and (f)
$v=-3t_y$  into the Brillouin zone of the original model,
respectively.   The area surrounded  by the  solid line is the
Brillouin zone of the original model. The yellow shaded area is the
image of the Brillouin zone of the modified model with a staggered
potential mapping into the Brillouin zone of the original model. The
filled circles represent the $\Upsilon$-invariant points in the
Brillouin zone of the original model. } \label{fig5}
\end{figure}

For the modified model with a staggered potential, the total
Hamiltonian violates the hidden symmetry, i.e., $[\Upsilon,H_1]\neq
0$.  However, Dirac points still exist and just  move to other
points in the Brillouin zone before the magnitude of the staggered
potential $v$ arrives at the critical value. Due to the von
Neumann-Wigner theorem, in the two-dimensional lattice, the band
degeneracy must be protected by a symmetry.\cite{vNW,Balents} After
the hidden symmetry $\Upsilon$ is violated, which symmetry protects
the band degeneracy at the Dirac points? In the following, we will
explain which symmetry is responsible for that.

Now we define a mapping as
\begin{eqnarray}
\Omega_v:(\bi{k}, \mathcal{H}_1(\bi{k}),
\Psi^{(1)}_\bi{k}(\bi{r}))\mapsto (\bi{K}, \mathcal{H}_0(\bi{K}),
\Psi^{(0)}_{\bi{K}}(\bi{r})),\label{trans}
\end{eqnarray}
which  maps the Bloch Hamiltonian (\ref{BH1}) into the form as
\begin{eqnarray}
{\cal H}_0(\bi{K}) &=&-2t_x'\cos K_x \sigma_x-2t_y' \cos K_y
\sigma_z, \label{BH10}
\end{eqnarray}
where  $t_x'=t_x$ and $t_y'=t_y+|v|/2$. Eq.(\ref{BH10}) is just the
Bloch Hamiltonian (\ref{BH0}) of the original model  except for the
different notations of the parameters. Eq.(\ref{BH10}) must have the
$\Upsilon$-invariant points in Brillouin zone as $M_{1,2}=( \pi/2,
\pm \pi/2)$  and $M_{3,4}=(0,\pm\pi/2)$  as shown in
Fig.\ref{fig2}(b).
 For the wave vectors, the transformation
 (\ref{trans})  is explicitly written as
 \begin{eqnarray}
 K_x &=&k_x,\  \mbox{for}\  k_x\in [-\pi/2, \pi/2],\label{KTX}\\
 K_y &=&\left\{\matrix{-\arccos\left(\frac{v+2t_y\cos k_y}{|v|+2t_y}\right), & k_y\in[-\pi, 0]\cr
 \arccos\left(\frac{v+2t_y\cos k_y}{|v|+2t_y}\right), & k_y\in[0,
 \pi]}\right.,\label{KTY}
 \end{eqnarray}
 and   for the Bloch functions, it can be written as
\begin{eqnarray}
\Omega_v  \Psi^{(1)}_\bi{k}(\bi{r})=\Psi^{(0)}_\bi{K}(\bi{r}),
\label{bmap}
\end{eqnarray}
where $\Psi^{(0)}_\bi{K}(\bi{r})$ is the Bloch function of the
original model. The shift from $\bi{k}$ to $\bi{K}$ after the
transformation is $\bi{K}-\bi{k}=(0,\delta_{k_y})$ with
 \begin{eqnarray}
 \delta_{k_y}=\left\{\matrix{-k_y-\arccos\left(\frac{v+2t_y\cos k_y}{|v|+2t_y}\right), & k_y\in[-\pi, 0]\cr
 -k_y+\arccos\left(\frac{v+2t_y\cos k_y}{|v|+2t_y}\right), & k_y\in[0,
 \pi]}\right.. \label{dely}
 \end{eqnarray}
The transformation (\ref{KTX}) implies that $k_x$ mapping into $K_x$ is just an equivalence. From Eq.(\ref{KTY}), we note that the mapping $k_y\rightarrow K_y$ is
 more complicated.
 When $v$ is positive, $\Omega_v$ maps the range  $[-\pi, \pi]$ for $k_y$   into  the range $ [-\arccos((v-2t_y)/(|v|+2t_y)), \arccos((v-2t_y)/(|v|+2t_y))] $ for $K_y$.
For the points  located  on the top and bottom boundaries of the
Brillouin zone of the modified model with a staggered potential, due
to the equivalence between points on the boundary of the Brillouin
zone,  the mapping is not one-to-one. For the interior points of the
Brillouin zone of the modified model with a staggered potential, the
mapping is continuous and one-to-one. On the whole, the mapping is
not surjective for non-vanishing $v$, i.e., the whole Brillouin zone
of the modified model with a staggered potential maps into part of
the Brillouin zone of the original model, as shown in
Figs.\ref{fig5}(a),(b) and (c). When $v$ is negative, $\Omega_v$
maps the ranges $[-\pi, 0]$ and $[0,\pi]$ for $k_y$ into the ranges
$[-\pi, -\arccos((v+2t_y)/(|v|+2t_y))]$ and
$[\arccos((v+2)/(|v|+2t_y)), \pi] $ for $K_y$, respectively.
Especially, $\Omega_v$ maps $k_y=0$   into
$K_y=\pm\arccos((v+2)/(|v|+2t_y))$, which is not one-to-one.
 In the interior part of each range, the mapping is  continuous and one-to-one. On the whole,
the  mapping is not surjective for non-vanishing $v$, as shown in
Figs.\ref{fig5}(d),(e) and (f).

We define a new hidden symmetry as
$\Lambda_v=\Omega_v^{-1}\Upsilon\Omega_v$, which consists of three
operators acting  in order on the wave vector and the Bloch
functions. Since
the operator $\Lambda_v$ depends on
$v$, the hidden symmetry evolves along with the magnitude of the staggered potential. When $v=0$, the operator $\Lambda_v$ returns to $\Upsilon$.
  For the wave vectors, the operation is performed as
$\Omega_v: \bi{k}\rightarrow \bi{K}$, $ \Upsilon: \bi{K}\rightarrow
\bi{K}'$ and $\Omega_v^{-1}:\bi{K}'\rightarrow \bi{k}'$. Considering
the explicit form of these transformations as
Eqs.(\ref{UT}),(\ref{KTX}) and (\ref{KTY}), we have
\begin{eqnarray}
\Lambda_v: \bi{k}= (k_x, k_y)\rightarrow \bi{k}'= (-k_x,
-k_y-\delta_{k_y}-\delta_{k'_y}+\pi). \label{symtrans}
\end{eqnarray}
 If the condition
$\bi{k}'=\bi{k}+\bi{K}_m$ is satisfied, then we can say that
$\bi{k}$ is a $\Lambda_v$-invariant point in the Brillouin zone of the modified model with a staggered potential.
Through Eqs.(\ref{dely}) and (\ref{symtrans}), we can show  that the
$\Lambda_v$-invariant points in the Brillouin zone of the modified model with
a staggered potential have the form as $P_{1,2}=(\pi/2,
\pm\arccos(-v/2t_y))$  and $P_{3,4}=(0,
\pm\arccos(-v/2t_y))$.  We  find that when $|v/t_y|=2$,
$\Lambda_v$-invariant points $P_1$ and $P_2$ are located
at the $k_y=0$ line or the top and bottom boundaries of the
Brillouin zone of the modified model with a staggered potential, so they meet together. That is so for $P_3$ and
$P_4$. However, when $|v/t_y|>2$, there is no solution for
$\Lambda_v$-invariant points, i.e. there does not exist any
$\Lambda_v$-invariant point.

For the square of the operator $\Lambda_v$, we have
 $\Lambda_v^2=\Omega_v^{-1}\Upsilon^2\Omega_v$. The hidden symmetry operator $\Lambda_v$
acting on the Bloch function $\Psi^{(1)}_\bi{k}(\bi{r})$ twice
successively has following the effect:
\begin{eqnarray}
\Lambda_v^2\Psi^{(1)}_\bi{k}(\bi{r})=e^{-2ik_x}\Psi^{(1)}_\bi{k}(\bi{r}),
\label{U3}
\end{eqnarray}
which can be derived from Eqs.(\ref{U2}), (\ref{KTX}), (\ref{KTY})
and (\ref{bmap}). Since $\Lambda_v$ is an antiunitary operator,
similar to Eq.(\ref{prot}), we have the following equation
\begin{eqnarray}
({\Psi^{(1)}_{P_i}}',\Psi^{(1)}_{P_i})&=&(\Lambda_v\Psi^{(1)}_{P_i},\Lambda_v{\Psi^{(1)}_{P_i}}')=(\Lambda_v\Psi^{(1)}_{P_i},\Lambda_v^2\Psi^{(1)}_{P_i})\nonumber\\
&=&e^{-2iP_{ix}}( {\Psi^{(1)}_{P_i}}', \Psi^{(1)}_{P_i}).
\end{eqnarray}
From Eq.(\ref{U3}), we have $\Lambda_v^2=-1$ at
$\Lambda_v$-invariant points $P_1$ and $P_2$, and
$\Lambda_v^2=1$ at $\Lambda_v$-invariant points $P_3$ and
$P_4$.
   Therefore, we have the solution $({\Psi^{(1)}_{P_{i}}},\Psi^{(1)}_{P_{i}})=0$ at the $\Lambda_v$-invariant points $P_1$ and $P_2$.
   We conclude  that the bands must be degenerate at the points $P_{1}$ and $P_2$
 but are not at the points $P_{3}$ and $P_4$, which is consistent with the dispersion relation calculated previously.
  That is to say, the Dirac points at $P_1$ and $P_2$ are
 protected by the hidden symmetry $\Lambda_v$. Since $\Lambda_v$ depends on the magnitude of the staggered potential $v$,
 the hidden symmetry protected degenerate points $P_1$ and $P_2$
evolve along with  changing  of the parameter $v$ for $|v/t_y|<2$.
For the case $|v/t_y|=2$, $P_1$ and $P_2$ are the same points, so
the Dirac points merge. When  $|v/t_y|>2$,   the
$\Lambda_v$-invariant points $P_1$ and $P_2$  do not exist, so a
gap opens.

 We can interpret the hidden symmetry protection in a more intuitive
 way. Although the  modified model with a staggered potential
 violates the hidden symmetry $\Upsilon$, the operator $\Omega_v$ can map the Bloch Hamiltonian
 (\ref{BH1}) into the Bloch Hamiltonian (\ref{BH0}), which is just the  Bloch Hamiltonian of the original model.
 However, the mapping $\Omega_v$  is not surjective. That is, the Brillouin zone of
the modified model with a staggered model
   maps into   part of the Brillouin zone of the original model
 as shown in Fig.\ref{fig5}.  When $|v/t_y|<2$, the image of the
 Brillouin zone of the modified model with a staggered model includes the
 $\Upsilon$-invariant points in the Brillouin zone of the original model
 as shown in Fig.\ref{fig5}(a) and (d). There always exist four points in the Brillouin zone of the modified model with a staggered model  mapping into
  the four $\Upsilon$-invariant points in the Brillouin zone of the original model.
    When $|v/t_y|=2$, the two
  points on the boundary or $k_y=0$ line of the Brillouin zone of the modified model with a staggered potential
  map into the four $\Upsilon$-invariant points in the Brillouin zone of the original model, as shown in Fig.\ref{fig5}(b) and (e).
  The two Dirac points meet  and merge.   When $|v/t_y|>2$, the
 Brillouin zone of the modified model with a staggered potential maps into part of the Brillouin zone of the original model, which does not include
 the $\Upsilon$-invariant points  as shown in Fig.\ref{fig5}(c) and (f).
 This is to say, there does not exist any point in the Brillouin
 zone of the modified model with a staggered potential that can map into the
 $\Upsilon$-invariant points in the Brillouin zone of the original model. Thus, there are no symmetries to support the existence fo  Dirac points.

\subsection{The modified model with  the diagonal hopping terms}

\begin{figure}[ht]
\includegraphics[width=0.25\columnwidth]{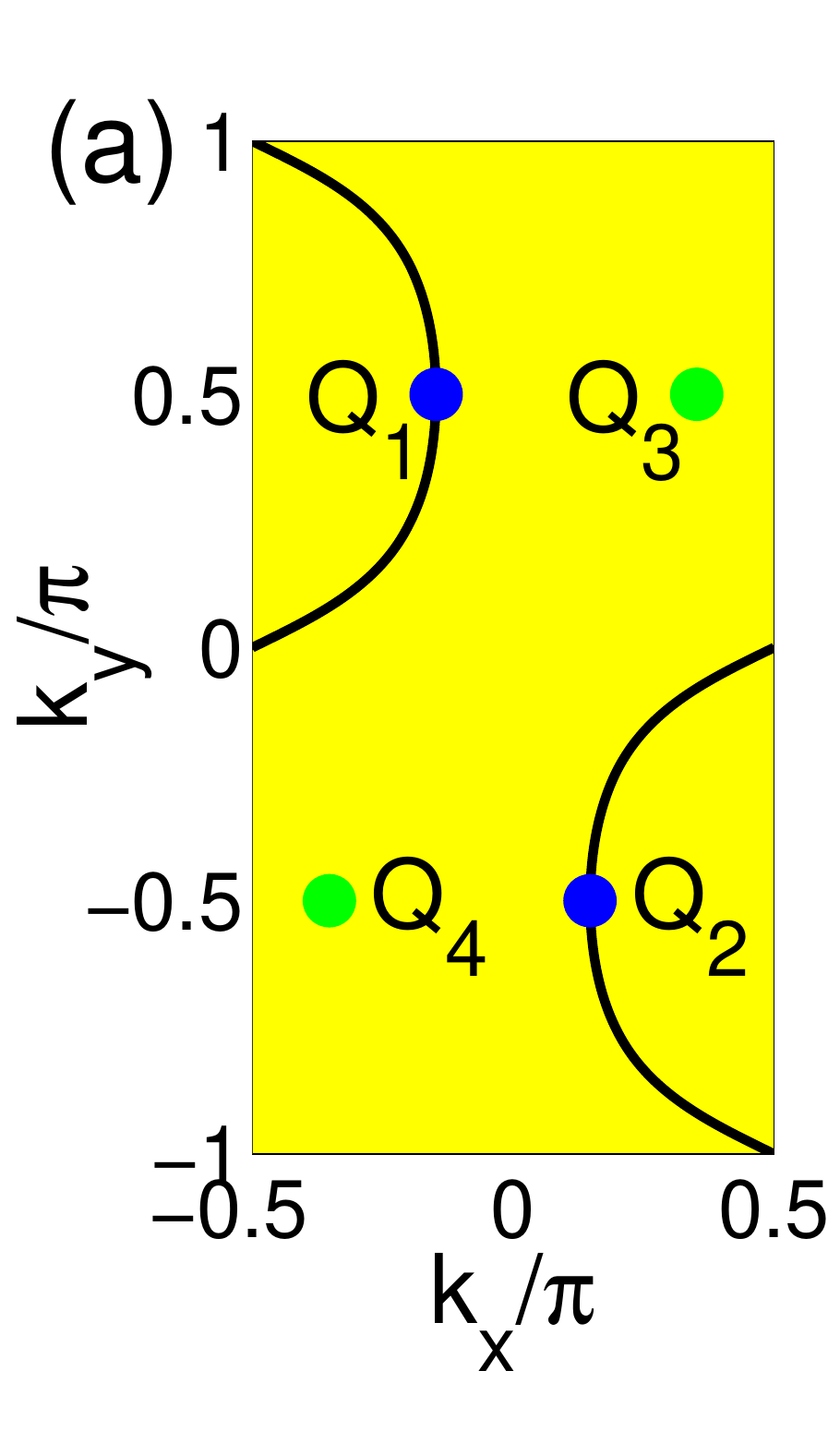}
\includegraphics[width=0.4286\columnwidth]{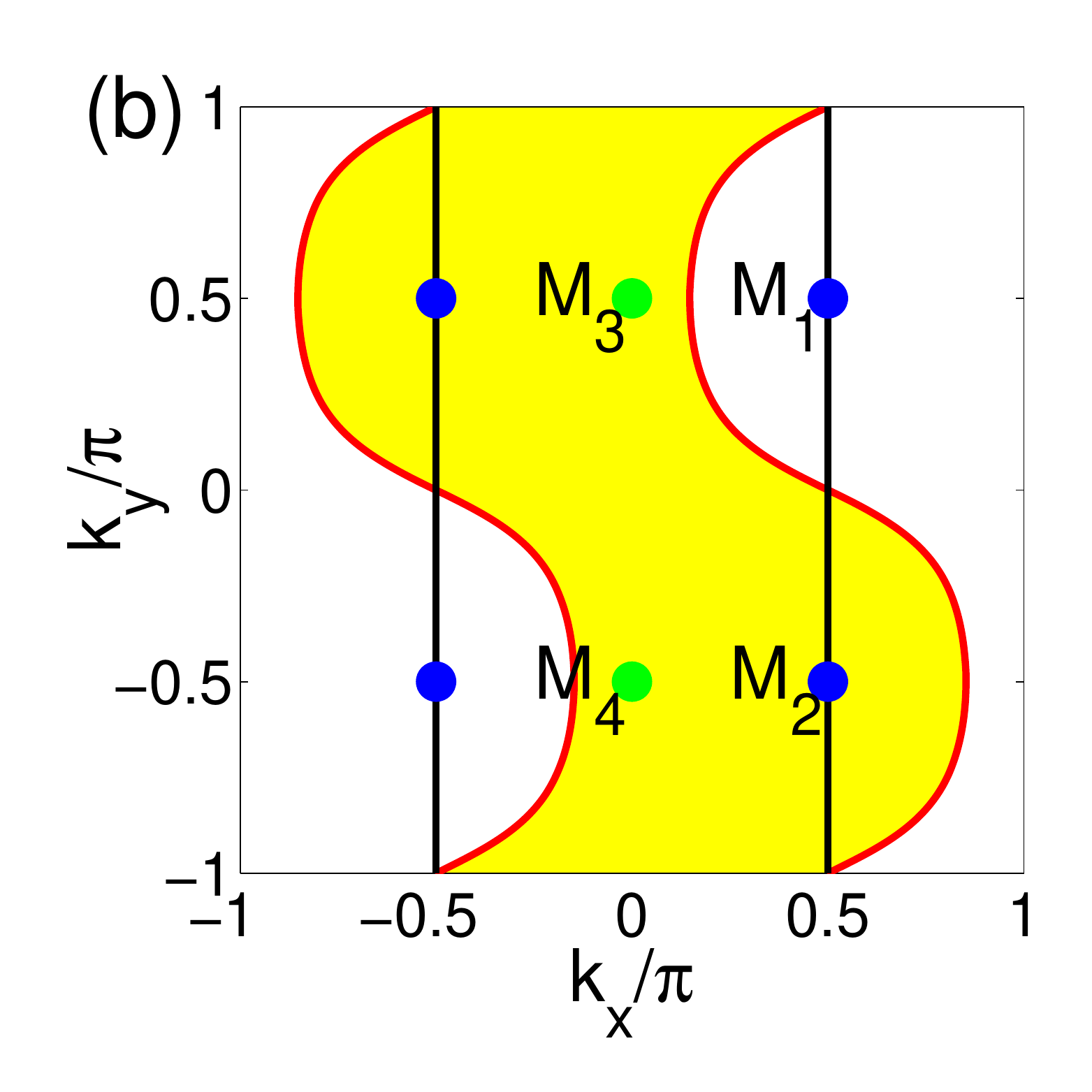}
\includegraphics[width=0.25\columnwidth]{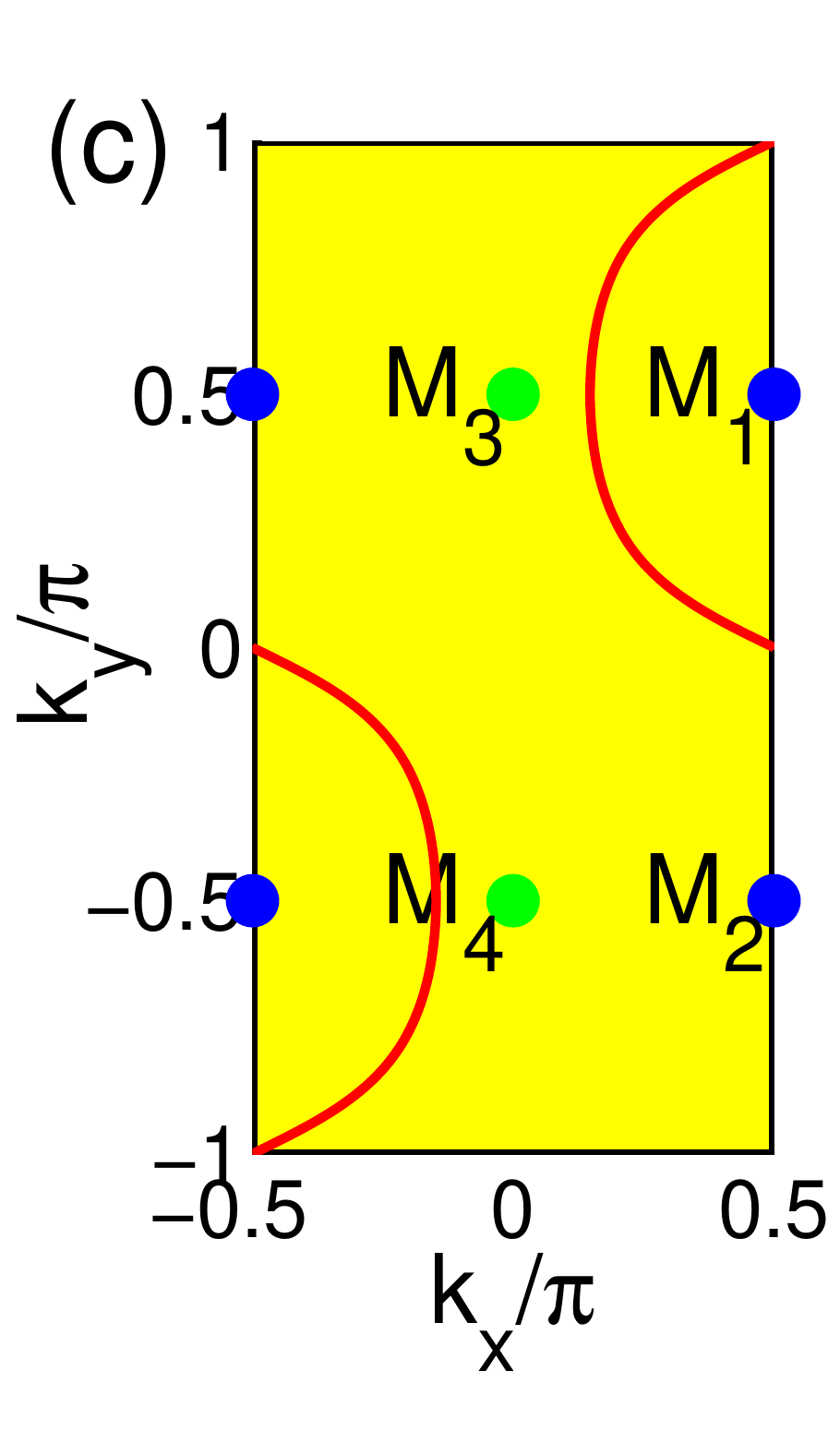}
\includegraphics[width=0.25\columnwidth]{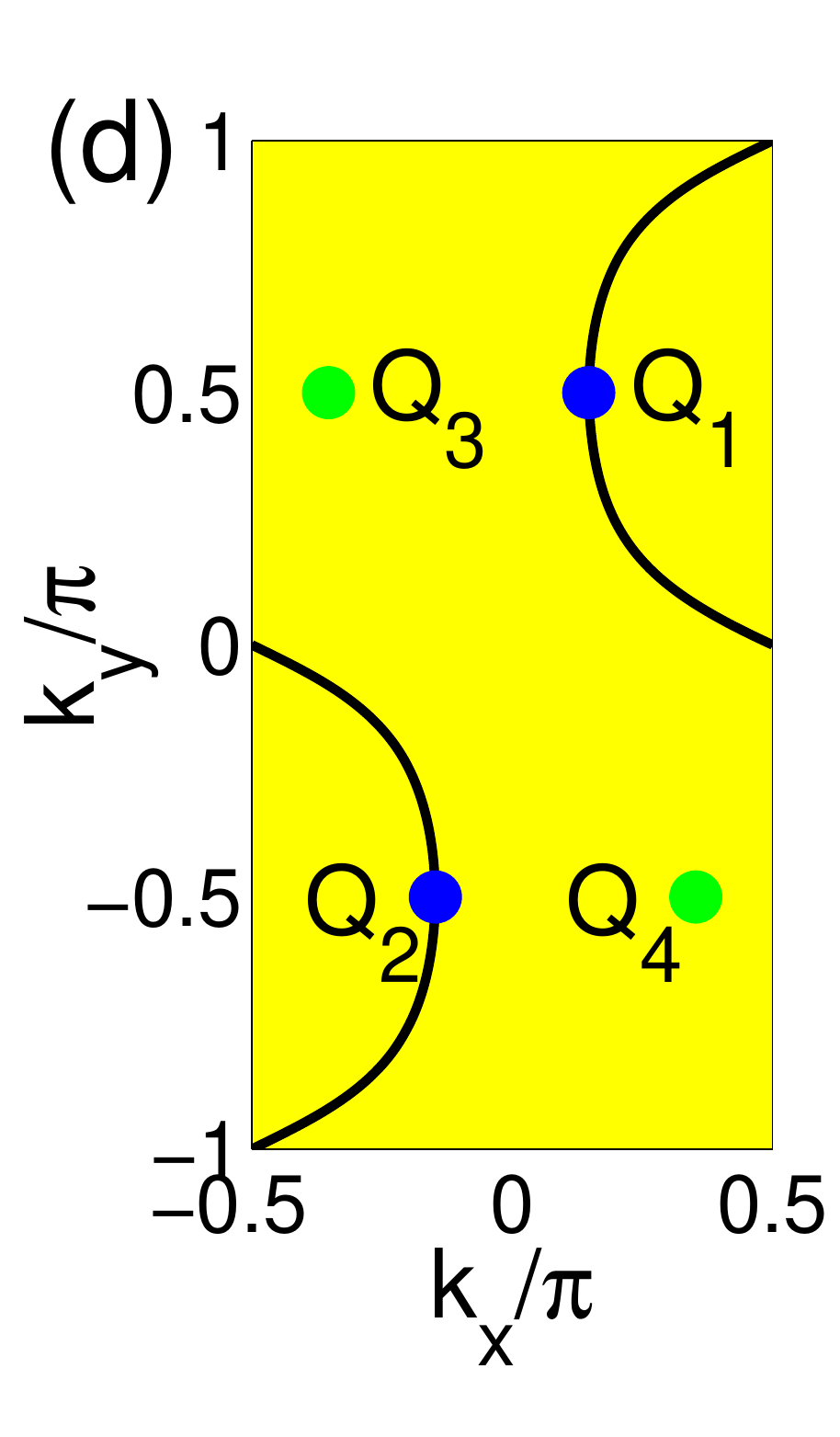}
\includegraphics[width=0.4286\columnwidth]{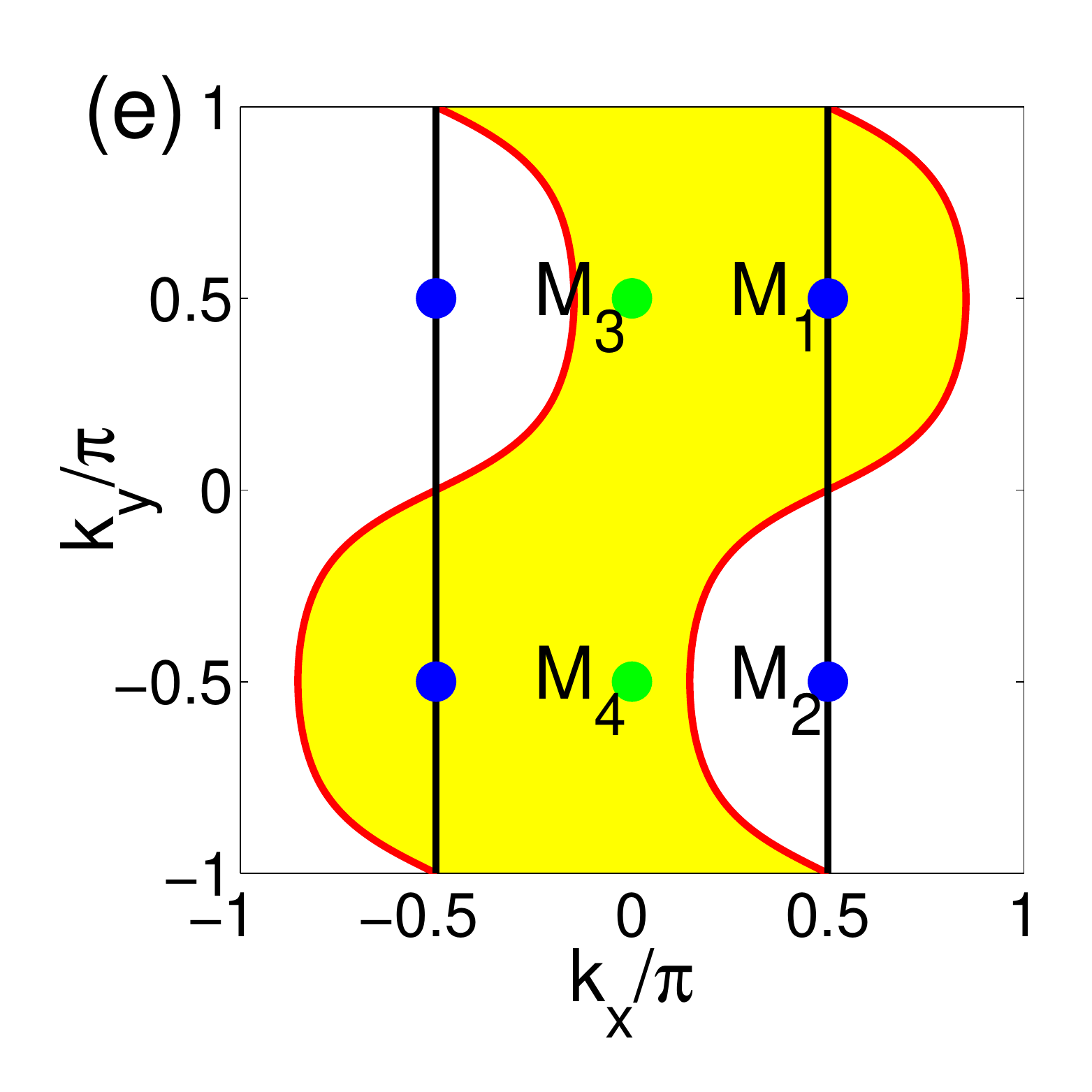}
\includegraphics[width=0.25\columnwidth]{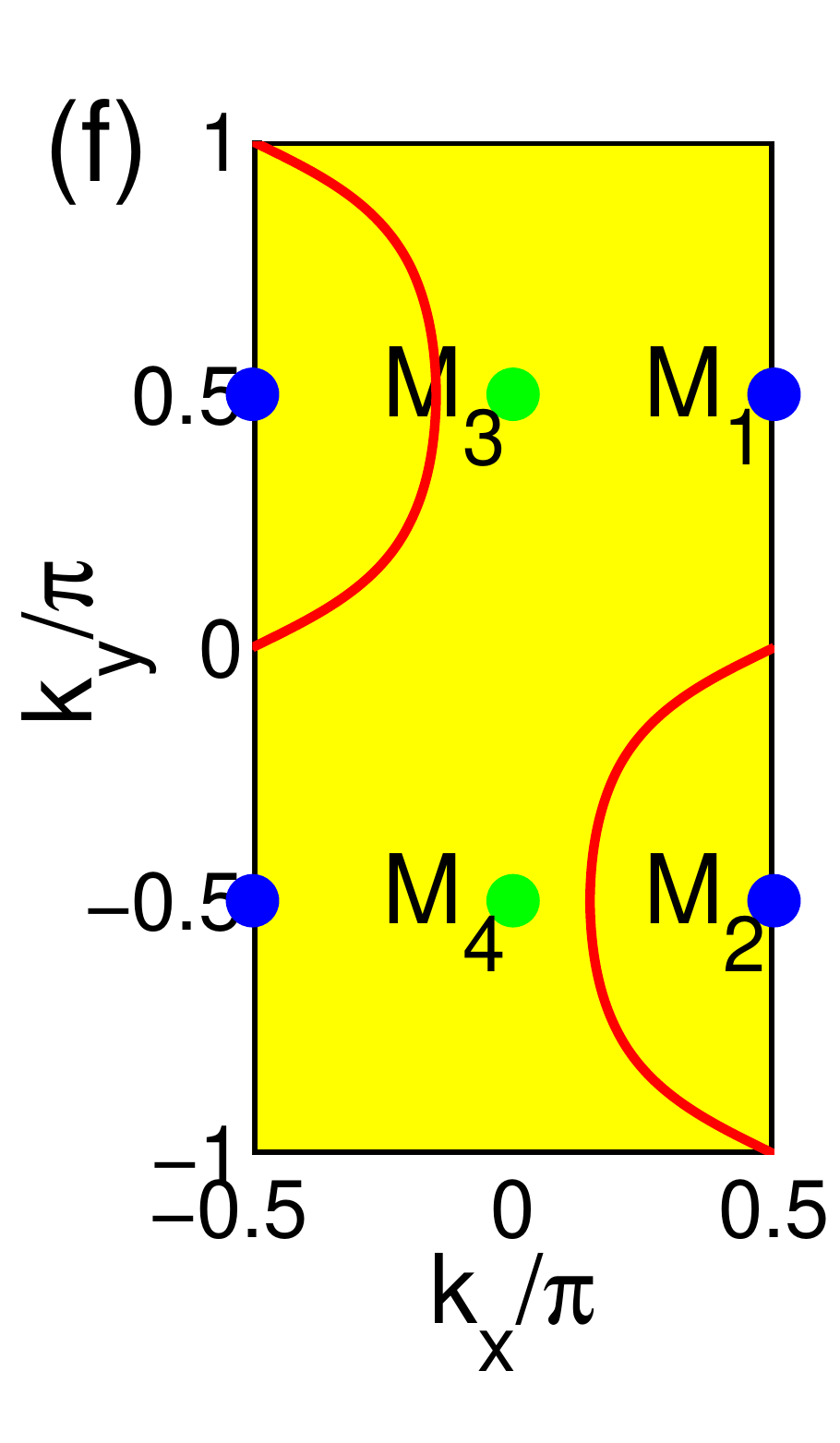}

\caption{(color online). The mapping of the Brillouin zone of the
modified model  with the diagonal hopping terms into the Brillouin
zone of the original model. (a) The Brillouin zone of the modified
model with the diagonal hopping terms,  (b) the image of mapping of
the modified model with the diagonal hopping terms in the momentum
space of the original model, and (c) the image of mapping of the
modified model with the diagonal hopping terms restricted in the
Brillouin zone of the original model  for $t_{xy}= t_x$; (d) the
Brillouin zone of the modified model with the diagonal hopping
terms, (e) the image of mapping of the modified model with the
diagonal hopping terms in the momentum space of the original model,
and (f) the image of mapping of the modified model with the diagonal
hopping terms restricted in the Brillouin zone of the original model
for $t_{xy}=-t_x$. Here, $Q_i$ denotes the
$\Lambda'_{t_{xy}}$-invariant points in the Brillouin zone of the
modified model with the diagonal hopping terms, and $M_i$ denotes
the $\Upsilon$-invariant points in the Brillouin zone of the
original model.  The black curves in (a) and (d)   map into the left
and right boundary of the Brillouin zone of the original model. }
\label{fig6}
\end{figure}

In the modified model with the diagonal hopping terms, the hidden
symmetry $\Upsilon$ respected by the original model is violated due
to the additive diagonal hopping terms. However, in this model, the
Dirac points do not vanish, so there must be some symmetry to
protect them.

Similarly, we define a mapping from the modified model with the
hopping terms to the original model as
\begin{eqnarray}
\Omega'_{t_{xy}}:(\bi{k}, \mathcal{H}_2(\bi{k}),
\Psi^{(2)}_\bi{k}(\bi{r}))\mapsto (\bi{K}, \mathcal{H}_0(\bi{K}),
\Psi^{(0)}_{\bi{K}}(\bi{r})),\label{map2}
\end{eqnarray}
where $\Psi^{(2)}_\bi{k}(\bi{r})$ is the Bloch function for the
modified model with the diagonal hopping terms. Toward that end, we
first rewrite the Bloch Hamiltonian (\ref{BH2}) as
\begin{eqnarray}
{\cal H}_2(\bi{k})
&=&-2T_{k_y}\cos(k_x-\alpha_{k_y})\sigma_x-2t_y\cos k_y\sigma_z
\end{eqnarray}
 where the parameter $T_{k_y}$ is defined as $T_{k_y}= \sqrt{t_x^2+4t_{xy}^2\sin^2k_y}$, which  depends on   $k_y$; the parameter $ \alpha_{k_y}$ also
 depends on $k_y$ and is defined by  $\alpha_{k_y}=\arctan(2t_{xy}\sin k_y/t_x)$. If we suppose that the
mapping $\Omega'_{t_{xy}}$ has the effect as
\begin{eqnarray}
\Omega'_{t_{xy}}: (k_x,k_y)\rightarrow
(K_x,K_y)=(k_x-\alpha_{k_y},k_y),\label{Theta}
\end{eqnarray}
and $T_{k_y}\rightarrow t_x'=T_{k_y}\cos\alpha_{k_y}=t_x$ and
$t_y\rightarrow t_y'=t_y$, then ${\cal H}_2(\bi{k})$ maps into
${\cal H}_0(\bi{K})$  as
\begin{eqnarray}
{\cal H}_0(\bi{K})&=&-2t_x'\cos K_x \sigma_x-2t_y'\cos K_y\sigma_z,
\end{eqnarray}
 which is just the Bloch Hamiltonian (\ref{BH0}) except for the
different notations of the parameters. For the Bloch functions, we
have
\begin{eqnarray}
\Omega'_{t_{xy}}\Psi^{(2)}_{\bi{k}}(\bi{r})=\Psi^{(0)}_{\bi{K}}(\bi{r}).
\end{eqnarray}

We define a new hidden symmetry as
$\Lambda'_{t_{xy}}={\Omega'}^{-1}_{t_{xy}}\Upsilon\Omega'_{t_{xy}}$.
For the wave vectors, the operation is performed as
$\Omega'_{t_{xy}}:\bi{k}\rightarrow \bi{K}$, $\Upsilon:
\bi{K}\rightarrow \bi{K}'$,
${\Omega'}_{t_{xy}}^{-1}:\bi{K}'\rightarrow \bi{k}'$. Considering
the explicit form of these transformations as Eqs.(\ref{UT}) and
(\ref{Theta}), we have
\begin{eqnarray}
\Lambda'_{t_{xy}}:
\bi{k}=(k_x,k_y)\rightarrow\bi{k}'=(-k_x+\alpha_{k_y}+\alpha_{k_y'},-k_y+\pi)
\end{eqnarray}
 If the condition
$\bi{k}'=\bi{k}+\bi{K}_m$ is satisfied, then we can say that
$\bi{k}$ is a $\Lambda'_{t_{xy}}$-invariant point in the Brillouin
zone of the modified model with the diagonal hopping terms. In this
model, the $\Lambda'_{t_{xy}}$-invariant points in the Brillouin
zone are $Q_1=(-\arctan(t_x/2t_{xy}),\pi/2)$,
$Q_2=(\arctan(t_x/2t_{xy}),-\pi/2)$,
$Q_3=(\arctan(2t_{xy}/t_x),\pi/2)$ and
$Q_4=(-\arctan(2t_{xy}/t_x),-\pi/2)$, as shown in Fig.\ref{fig6} (a)
for the case of $t_{xy}/t_x>0$ and Fig.\ref{fig6}(d) for the case of
$t_{xy}/t_x<0$.

For the square of the operator $\Lambda'_{t_{xy}}$, we have
${\Lambda'}_{t_{xy}}^2={\Omega'}_{t_{xy}}^{-1}\Upsilon^2\Omega'_{t_{xy}}={\Omega'}_{t_{xy}}^{-1}T_{\bi{a}_1}\Omega'_{t_{xy}}$.
The hidden symmetry operator $\Lambda'_{t_{xy}}$ acting on the Bloch
function $\Psi^{(2)}_{\bi{k}}(\bi{r})$ twice successively has the
effect as
\begin{eqnarray}
{\Lambda'}_{t_{xy}}^2
\Psi^{(2)}_{\bi{k}}(\bi{r})=e^{-2i(k_x-\alpha_{k_y})}\Psi^{(2)}_{\bi{k}}(\bi{r})\label{lam2}
\end{eqnarray}
which  can be derived from Eqs.(\ref{U2}) and (\ref{Theta}).
 Since $\Lambda'_{t_{xy}}$ is an antiunitary operator,
similar to Eq.(\ref{prot}), we have the following equation
\begin{eqnarray}
 ({\Psi^{(2)}_{Q_i}}',\Psi^{(2)}_{Q_i})&=&(\Lambda'_{t_{xy}}\Psi^{(2)}_{Q_i},\Lambda'_{t_{xy}}{\Psi^{(2)}_{Q_i}}')
=
(\Lambda'_{t_{xy}}\Psi^{(2)}_{Q_i},{\Lambda'_{t_{xy}}}^2\Psi^{(2)}_{Q_i})
\nonumber\\&=&e^{-2i(Q_{ix}-\alpha_{Q_{iy}})}( {\Psi^{(2)}_{Q_i}}',
\Psi^{(2)}_{Q_i}).
\end{eqnarray}
 From
Eq.(\ref{lam2}), we can obtain ${\Lambda'}^2_{t_{xy}}=-1$ at the
$\Lambda'_{t_{xy}}$-invariant points $Q_1$ and $Q_2$, and
${\Lambda'}^2_{t_{xy}}=1$ at the $\Lambda'_{t_{xy}}$-invariant
points $Q_3$ and $Q_4$.
 Therefore, we have the solution $({\Psi^{(2)}_{Q_{i}}}',\Psi^{(2)}_{Q_{i}})=0$ at the $\Lambda'_{t_{xy}}$-invariant points $Q_1$ and $Q_2$.
   We conclude  that the bands must be degenerate at the points $Q_{1}$ and $Q_2$
 while the band degeneracy is not guaranteed at the points
$Q_3$ and $Q_4$, which is consistent with the dispersion relation calculated previously.
  That is to say, the Dirac points at $Q_1$ and $Q_2$ are
 protected by the hidden symmetry $\Lambda'_{t_{xy}}$.
 The hidden symmetry $\Lambda'_{t_{xy}}$
evolves along with the parameter $t_{xy}$. It is easy to find that
when $t_{xy}=0$, the hidden symmetry operator $\Lambda'_{t_{xy}}$
returns to the operator $\Upsilon$ and the degenerate points $Q_1$
and $Q_2$ are just the points $M_1$ and $M_2$. When $t_{xy}$
changes, the degenerate points $Q_1$ and $Q_2$ move towards opposite
directions, respectively. When $t_{xy}$ approaches   infinity, the
degenerate points approach   the $k_y=0$ line from two sides,
respectively. For any value of the parameter $t_{xy}$, the
degenerate points $Q_1$ and $Q_2$ do not merge and no gap opens. All
these conclusions are consistent with the dispersion relation
calculated previously.

We can interpret the above conclusions from the mapping of the
Brillouin zone of the modified model with the diagonal hopping terms
 to the Brillouin zone of the original model, which is shown in
Figs.\ref{fig6}(a),(b),(c) for the case of $t_{xy}/t_x>0$ and
Figs.\ref{fig6}(d),(e),(f) for the case of $t_{xy}/t_x<0$.
Figs.\ref{fig6}(a) and (d) show the Brillouin zone of the modified
model with the diagonal hopping terms. Figs.\ref{fig6}(b) and (e)
show the image of the mapping $\Omega'_{t_{xy}}$ of the Brillouin
zone of the modified model with the diagonal hopping terms in the
momentum space of the original model. If the image of the mapping
$\Omega'_{t_{xy}}$ is restricted in the Brillouin zone of the
original model, it is like that   shown in Figs.\ref{fig6} (c) and
(f). It is easy to find that the mapping $\Omega'_{t_{xy}}$
 just shifts
the points in the Brillouin zone along the $x$ direction as shown in
Figs.\ref{fig6}(b) and (e).  The mapping $\Omega'_{t_{xy}}$   is
one-to-one and  surjective, which can be found from
Figs.\ref{fig6}(c) and (d). Specifically,  the left and right
boundaries of the Brillouin zone of the modified model with the
diagonal hopping terms as shown in Figs.\ref{fig6} (a) and (b) map
into the red solid lines in the Brillouin zone of the original model
as shown in Figs.\ref{fig6}(c) and (f). The black curved lines in
the Brillouin zone of the modified model with the diagonal hopping
terms as shown in Figs.\ref{fig6} (a) and (b) map into the left and
right boundary of the Brillouin zone of the original model, where
the $\Upsilon$-invariant points $M_1$ and $M_2$ are located.
 The $\Lambda'_{t_{xy}}$-invariant points
$Q_i (i=1,2,3,4)$ in the Brillouin zone of the modified model with
the diagonal hopping terms map into the $\Upsilon$-invariant points
$M_i (i=1,2,3,4)$ in  the Brillouin zone of the original model.
Since the mapping is surjective, there always exist points  $Q_1$
and $Q_2$ in the Brillouin zone of the modified model with the
diagonal hopping terms mapping into the $\Upsilon$-invariant points
$M_1$ and $M_2$ in the Brillouin zone of the original model.
Therefore, the Dirac points always are protected by a hidden
symmetry and no gap opens for any value of the parameter $t_{xy}$.
Because the corresponding hidden symmetry $\Lambda'_{t_{xy}}$
evolves along with the parameter $t_{xy}$, the Dirac points move as
the parameter $t_{xy}$ changes.

 \section{Conclusion}

In summary, we have studied the original model, a fermionic square
lattice with only the horizontal and vertical hopping terms, and the
two modified models with a staggered potential and the diagonal
hopping terms, respectively.   All   three models support the
existence of  massless Dirac fermions. In the original model, there
are two Dirac points in the Brillouin zone, which are protected by a
hidden symmetry. In the modified model with a staggered potential,
the  two Dirac points move away from or approach  each other with
increasing of the  magnitude of the staggered potential. When the
magnitude arrives at a critical value, the two Dirac points merge at
the $k_y=0$ line or the $k_y=\pi$ line which is determined by the
sign of the staggered potential. When the magnitude  of the
staggered potential is greater than the critical value, a gap opens,
and the system becomes an insulator. In the modified model with the
diagonal hopping terms, the two Dirac points in the Brillouin move
with increasing   amplitude of the diagonal hopping in
   two opposite directions, respectively.
      But the Dirac points never vanish and the system is always gapless  for any amplitude of the diagonal hopping.
For the two modified models, we have developed a mapping method that
maps the modified models into the original model, to find hidden
symmetries evolving with the parameters. The moving of the Dirac
points in the Brillouin zone for the two modified models can be
explained by the evolution of the hidden symmetries along with the
parameters. The merging of Dirac points in  the modified model with
a staggered potential can also be explained by the disappearance of
the hidden-symmetry-invariant points in the Brillouin zone when the
parameter is beyond the critical value.
 The original model can  be realized experimentally and
detected in an optical lattice  with laser-assisted tunneling as
proposed in Reference [\onlinecite{Hou2}]. Based on the original
model, two modified models can also be realized with the existing
techniques on optical lattices. The topological charge at Dirac
points can be detected by the interferometric approach.\cite{Abanin}

\begin{acknowledgments}
We thank W. Chen  for helpful discussions. This work was  supported by the National Natural Science Foundation
of China under Grants No. 11274061 and No. 11004028.
\end{acknowledgments}


\begin{thebibliography}{99}

\bibitem{Novoselov1} K.S. Novoselov, A.K. Geim, S.V. Morozov, D. Jiang, Y.
Zhang, S.V. Dubonos, I.V. Grigorieva, and A. A. Firsov, Science
\textbf{306}, 666 (2004).

\bibitem{Novoselov2} K.S. Novoselov, A.K. Geim, S.V. Morozov, D. Jiang, M.I. Katsnelson,
 I.V. Grigorieva, S.V. Dubonos, and A.A. Firsov, Nature (London) \textbf{438}, 197
 (2005). 

\bibitem{Zhang} Y. Zhang, Y.W. Tan, H.L. Stormer,  and Philip Kim, Nature (London) \textbf{438}, 201 (2005).

\bibitem{Gusynin} V. P. Gusynin and S. G. Sharapov, Phys. Rev. Lett. \textbf{95}, 146801
(2005). 

\bibitem{Li} G. Li and E. Y. Andrei, Nat. Phys. \textbf{3}, 623 (2007). 

\bibitem{Hou2} J.M. Hou, W.X. Yang and X.J. Liu, Phys. Rev. A \textbf{79}, 043621 (2009).

\bibitem{Zhu} S.L. Zhu, B. Wang, and L.M. Duan, Phys. Rev. Lett. \textbf{98}, 260402 (2007).

\bibitem{Goldman}N. Goldman, A. Kubasiak, A. Bermudez, P. Gaspard, M. Lewenstein, and M. A. Martin-Delgado,
Phys. Rev. Lett. \textbf{103}, 035301 (2009). 

\bibitem{Bercioux}D. Bercioux, D. F. Urban, H. Grabert, and W. H\"ausler,
Phys. Rev. A \textbf{80}, 063603 (2009). 

\bibitem{Goldman2}N. Goldman, E. Anisimovas, F. Gerbier, P. \"Ohberg, I. B. Spielman, G.
Juzeli\={u}nas, New J. Phys. \textbf{15}, 013025 (2013).

\bibitem{Tarruell} L. Tarruell, D. Greif, T. Uehlinger, G. Jotzu, and  T. Esslinger, Nature (London) \textbf{483}, 302
 (2012). 

\bibitem{Fu} L. Fu, C.L. Kane, and E.J. Mele, Phys. Rev. Lett. \textbf{98}, 106803 (2007).

\bibitem{Moore} J.E. Moore and L. Balents, Phys. Rev. B \textbf{75}, 121306 (2007).

\bibitem{Roy} R. Roy, Phys. Rev. B \textbf{79}, 195322 (2009). 

\bibitem{Wan} X. Wan, A.M. Turner, A. Vishwanath, and S.Y. Savrasov,
Phys. Rev. B \textbf{83}, 205101 (2011). 

\bibitem{Xu} G. Xu, H. Weng, Z. Wang, X. Dai, and Z. Fang, Phys.
Rev. Lett. \textbf{107}, 186806 (2011). 

\bibitem{Burkov} A.A. Burkov and L. Balents, Phys. Rev. Lett.
\textbf{107}, 127205 (2011). 

\bibitem{Jiang} J.H. Jiang, Phys. Rev. A \textbf{85}, 033640 (2012).

\bibitem{Delplace} P. Delplace, J. Li, and D Carpentier, Europhys.
Lett. \textbf{97}, 67004 (2012). 

\bibitem{Hou1} J.M. Hou, Phys. Rev. Lett. \textbf{111}, 130403 (2013).

\bibitem{vNW} J. von Neumann and E. Wigner, Z. Phys. \textbf{30}, 467 (1929).

\bibitem{Balents} L. Balents, Physics \textbf{4}, 36 (2011). 

\bibitem{Volovik} G.E. Volovik, Lect. Notes in Phys. \textbf{870}, 343
(2013).

\bibitem{Abanin} D.A. Abanin, T. Kitagawa, I. Bloch, and E. Demler,
Phys. Rev. Lett. \textbf{110}, 165304 (2013). 

\end{thebibliography}
\end{document}